\documentclass[twocolumn,pre,amsmath,amssymb]{revtex4-1}

\usepackage{graphicx}
\usepackage{dcolumn}
\usepackage{bm}
\usepackage[table]{xcolor}
\usepackage[normalem]{ulem} 
\usepackage{hyperref}
\hypersetup{
    colorlinks = true,
    linkcolor = {blue},
    citecolor = {blue},
    urlcolor = {blue},
}

\newtheorem{proposition}{Proposition}

\newcommand{\bx}{\ensuremath{\boldsymbol{x}}}
\newcommand{\by}{\ensuremath{\boldsymbol{y}}}
\newcommand{\bp}{\ensuremath{\boldsymbol{p}}}
\newcommand{\bq}{\ensuremath{\boldsymbol{q}}}
\newcommand{\bbp}{\ensuremath{\boldsymbol{\bar{p}}}}
\newcommand{\bbq}{\ensuremath{\boldsymbol{\bar{q}}}}

\newcommand{\bbx}{\ensuremath{\boldsymbol{\bar{x}}}}

\newcommand{\bc}{\ensuremath{\boldsymbol{c}}}
\newcommand{\bd}{\ensuremath{\boldsymbol{d}}}

\newcommand{\dggm}{\ensuremath{Dg_{(m)}^{(3)}}}

\newcommand{\xmt}{\ensuremath{X_{(m)}^\tau}}
\newcommand{\ymt}{\ensuremath{Y_{(m)}^\tau}}

\newcommand{\rst}{\cellcolor[RGB]{240, 153, 149}}
\newcommand{\rnd}{\cellcolor[RGB]{249, 207, 205}}

\newcommand{\st}{\shortstack}

\begin{document}

\title{Topological time-series analysis with delay-variant embedding}
\author{Quoc Hoan Tran}
\email{zoro@biom.t.u-tokyo.ac.jp}

\affiliation{
	Department of Information and Communication Engineering, Graduate School of Information Science and Technology,\\
	The University of Tokyo, Tokyo 113-8656, Japan
}
\author{Yoshihiko Hasegawa}

\email{hasegawa@biom.t.u-tokyo.ac.jp}

\affiliation{
	Department of Information and Communication Engineering, Graduate School of Information Science and Technology,\\
	The University of Tokyo, Tokyo 113-8656, Japan
}

\date{\today}

\begin{abstract}
Identifying the qualitative changes in time-series data provides insights into the dynamics associated with such data. 
Such qualitative changes can be detected through topological approaches, which first embed the data into a high-dimensional space using a time-delay parameter and subsequently extract topological features describing the shape of the data from the embedded points.
However, the essential topological features that are extracted using a single time delay  
are considered to be insufficient for evaluating the aforementioned qualitative changes,
even when a well-selected time delay is used.
We therefore propose a delay-variant embedding method that constructs the extended topological features by considering the time delay as a variable parameter
instead of considering it as a single fixed value.
This delay-variant embedding method reveals multiple-time-scale patterns in a time series by allowing the observation of the variations in topological features, with the time delay serving as an additional dimension in the topological feature space.
We theoretically prove that the constructed topological features are robust
when the time series is perturbed by noise. 
Furthermore, we combine these features with the kernel technique in machine learning algorithms to classify the general time-series data. 
We demonstrate the effectiveness of our method for classifying the synthetic noisy biological and real time-series data. Our method outperforms a method
that is based on a single time delay and, surprisingly, achieves the highest classification accuracy on an average among the standard time-series analysis techniques.
\end{abstract}

\pacs{Valid PACS appear here}

\maketitle

\section{Introduction}
Time-series data can undergo qualitative changes such as transitioning 
from the quiescent state to oscillatory dynamics through a bifurcation. 
Identifying such changes enables deep understanding of the underlying dynamics; 
however, this is challenging when the data are subject to noise. 
In material science, topological features, which indicate the ``shape'' of the data,
can be used to detect the qualitative changes, i.e., phase transitions~\cite{pre:donato:phase:16,kusano:gskernel:2016} or transitions in morphological and hierarchical structures~\cite{pre:ardanza:granular:14,nakamura:nano:2015,hiraoka:pnas:2016,pre:ichinomiya:craze:17}.
Because topology is a qualitative property that is stable under the influence of noise,
the topological features of time-series data are expected to reflect the qualitative changes in dynamics. 
These features are constructed through delay embedding,
in which a time series $x(t)$ is mapped to $m$-dimensional points using 
delay coordinates $[x(t),x(t-\tau),...,x(t-(m-1)\tau)]$ on the embedded space,
where $\tau$ denotes the predefined time delay and $m$ denotes the embedding dimension.
Further, the embedded points form geometric features, such as clusters and loops,
and the topological features, which monitor the emergence and disappearance of geometric features, can be used to characterize the dynamics of the system~\cite{maletic:dynamic:2016,mittal:bifucation:2017}. 

Theoretically, for a noise-free time series of unlimited length,
embeddings with the same value of $m$ but different values of time delay $\tau$ are 
considered to be equivalent in terms of ``optimal" reconstruction~\cite{taken:1981}.
Herein, the reconstruction that preserves the invariants of dynamics, 
such as the fractal dimension and the Lyapunov exponent,
when the time series is embedded can be referred to as optimal reconstruction.
However, because a real time series is noisy and because of finite length,
the selection of $\tau$ is considered to be an inherently difficult problem.
Instead of being based on mathematically rigorous criteria, 
majority of the methods that are required for the selection of $\tau$
are observed to be based on heuristics, i.e.,
they involve a tradeoff between \textit{redundance} and \textit{irrelevance} 
in case of successive delay coordinates~\cite{casdagli:state:1991}.
Traditional methods determine $\tau$ as the first time scale
that maximizes the indices of independence 
such as linear independence and nonlinear independent information~\cite{fraser:pra:1986}, 
and the correlation sum~\cite{liebert:proper:1989}.
Meanwhile, the geometry-based strategies that are used for estimating $\tau$ 
examine the measures that are required for expanding 
the attractor in the reconstructed phase space~\cite{buzug:optimal:1992,buzug:comp:1992,rosenstein:recons:1994}.
However, the optimal value of $\tau$ exhibits no definite theoretical properties
because $\tau$ has no theoretical relevance in the mathematical framework of delay embedding~\cite{kantz:nonlinear:2003}.
Furthermore, there is no universal strategy for selecting $\tau$, 
and the value of $\tau$ that works well for one application 
may not work well for another application
even though the same data may be used~\cite{bradley:nonlinear:2015}.
Therefore, the optimal selection of $\tau$ depends on the objective of the analysis 
and is obtained by trial and error without using any systematic method.
This defect limits the power of 
topological features in time-series data analysis.

Herein, we propose a \emph{delay-variant embedding} method
in which the topological features are constructed
by considering $\tau$ as the variable parameter;
using this method, the topological changes are monitored in the embedded space.
Embedding with a single value of $\tau$ is sensitive to noise;
further, considering a range of $\tau$ can provide useful information
with which the dynamics can be understood.
We theoretically prove the stability of constructed features against noise and apply these features to classification of several time-series datasets. 
We demonstrate that our method outperforms a method that is based on a single value of $\tau$ 
in classifying the oscillatory activity of synthetic noisy biological data.
Surprisingly, in classifying real time-series data, 
our approach demonstrates a higher accuracy on an average
when compared to several standard techniques
that are used for performing time-series analysis.

\section{Method}
\subsection{Topological features from time-series data with delay-variant embedding}
To qualitatively evaluate the characteristics of the time series, 
we apply topological data analysis~\cite{carlsson:topology:2009}, 
which is a computational method that can be used 
for characterizing the topological features of high-dimensional data. 
We construct a simplicial-complex model~\cite{edels:topobook:2010} from the points 
in the embedded space and obtain the topological information as the number, position, and size of 
single- or multi-dimensional clusters and loops.
We build a complex over a set of points if the pairwise distances between them are less than or equal to $2\varepsilon$, where $\varepsilon$ is a given non-negative scale parameter. 
Different values of $\varepsilon$ result in different complexes and different topological information.
If $\varepsilon$ is considerably small, no connections are created;
further, the resulting simplicial complex is not different when compared to the original points.
As we gradually increase $\varepsilon$, connections appear between the points;
however, if $\varepsilon$ becomes considerably large, 
all the points are connected with each other,
and no useful information can be conveyed.
If we increase $\varepsilon$ as $\varepsilon_1 < \ldots < \varepsilon_n$, 
we obtain a sequence of embedded simplicial complexes 
that can be referred to as \emph{filtration} (see Appendix~\ref{sec:topo:features}). 

We use persistent homology theory~\cite{edelsbrunner:2002,zomorodian:2005} 
to study the topological features across filtration.
A practical way to visualize the results of persistent homology
is through multi-set points in the two-dimensional \emph{persistence diagram}.
In this diagram, each point $(b,d)$ represents an $l$-dimensional hole 
(i.e., the connected components are zero-dimensional, loops and tunnels are one-dimensional holes, and voids 
are two-dimensional holes) that appears at $\varepsilon=b$ (known as the \emph{birth scale}) and disappears at $\varepsilon=d$ (known as the \emph{death scale}) across the filtration (see Appendix~\ref{sec:topo:features}).
This information encapsulates the topological features of the time series after 
embedding and provides valuable insights into the behavior of a dynamical system;
for instance, the emergence of an oscillation in a time series can be attributed
to the birth of a loop in the embedded space.
Using these features, the qualitative properties of the time series 
can be captured in a robust and efficient manner. 

\begin{figure*}
	\includegraphics[width=15cm]{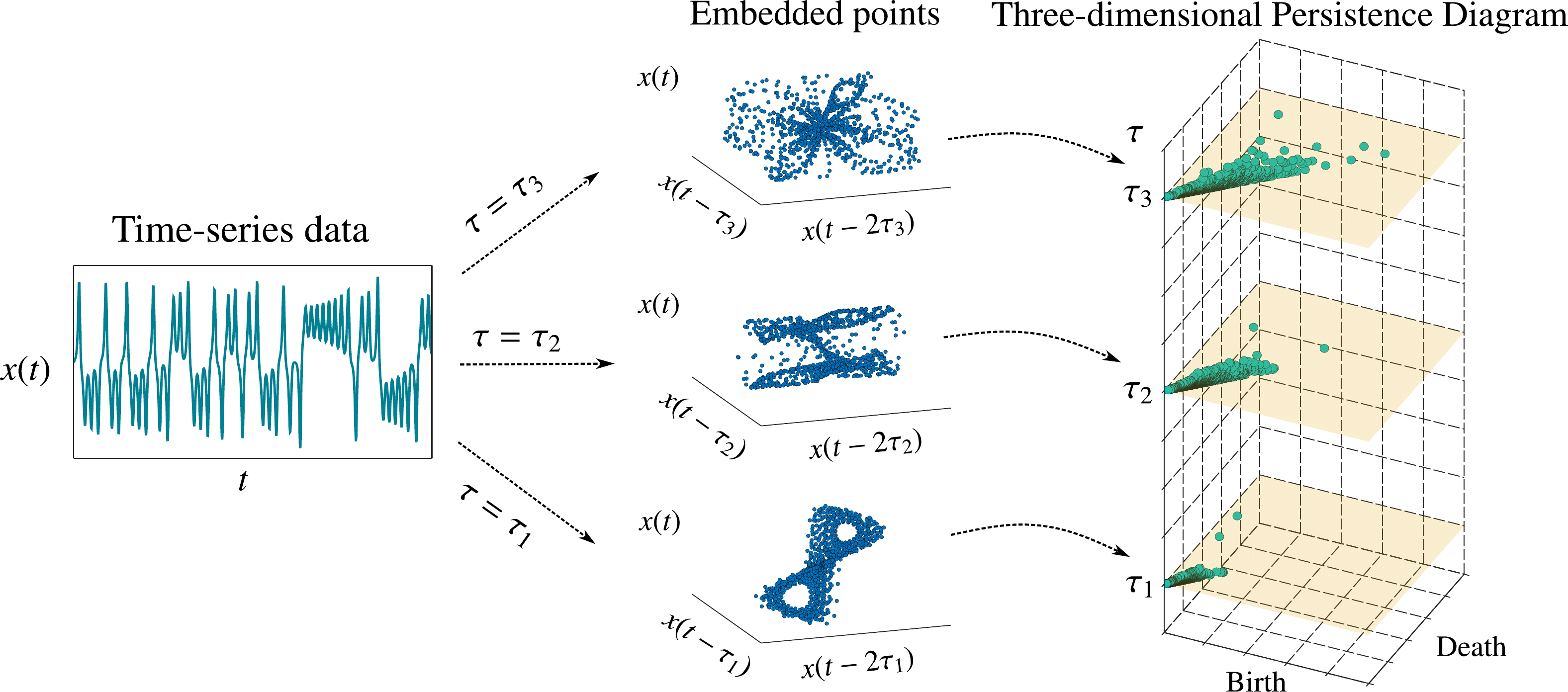}
	\protect\caption{Time series $x(t)$ is 
		embedded in an $m$-dimensional space (in this illustration, $m=3$) through delay embedding (left panel). 
		The embedded points at different values of time delay $\tau$ have different geometric features,
        such as clusters and loops (middle panel). 
		We extract the topological features such as the emergence (birth scale) and disappearance (death scale) of these geometric features.
		The birth and death scales at each $\tau$ are represented as points in a two-dimensional persistence diagram. 
		By observing the manner in which the topological features vary with $\tau$ serving as an additional dimension, 
		we obtain the three-dimensional persistence diagram (right panel),
		which can be considered to be a typical feature of the time series.
		\label{fig:overview_image}}
\end{figure*}

For the observed time series $x(t)$, we use $F_{l,\tau}(x(t))$ to denote the two-dimensional persistence diagram calculated for $l$-dimensional holes from the embedded points with time delay $\tau$.
We consider $\tau$ in a predefined set $\mathcal{T}=\{\tau_1,\tau_2,...,\tau_K\}$, where $\tau_1<\tau_2<\cdots<\tau_K$ ($K$ denotes the size of the set).
The three-dimensional persistence diagram for time series $x(t)$ can be defined as 
\begin{align}
    \text{PD}^{(3)}_{l}&(x(t))\nonumber\\
    &=\left \{(b,d,\tau)\mid (b,d) \in F_{l,\tau}(x(t)), \tau \in \mathcal{T}\right \}.
\end{align}
Figure~\ref{fig:overview_image} shows a schematic 
of the three-dimensional persistence diagram $\text{PD}^{(3)}_{1}(x(t))$ for loops and tunnels (one-dimensional holes), wherein the embedded points exhibit different shapes and topological features for different values of $\tau$.
In the middle panel of Fig.~\ref{fig:overview_image},
there are two big loops for $\tau_1$, whereas loops survive for short times with 
different distributions of birth and death scales (right panel) for $\tau_2$ and $\tau_3$.
This information cannot be obtained using a single value of $\tau$ in the embedding. 

\subsection{Stability of topological features}
\begin{figure}
	\includegraphics[width=8.5cm]{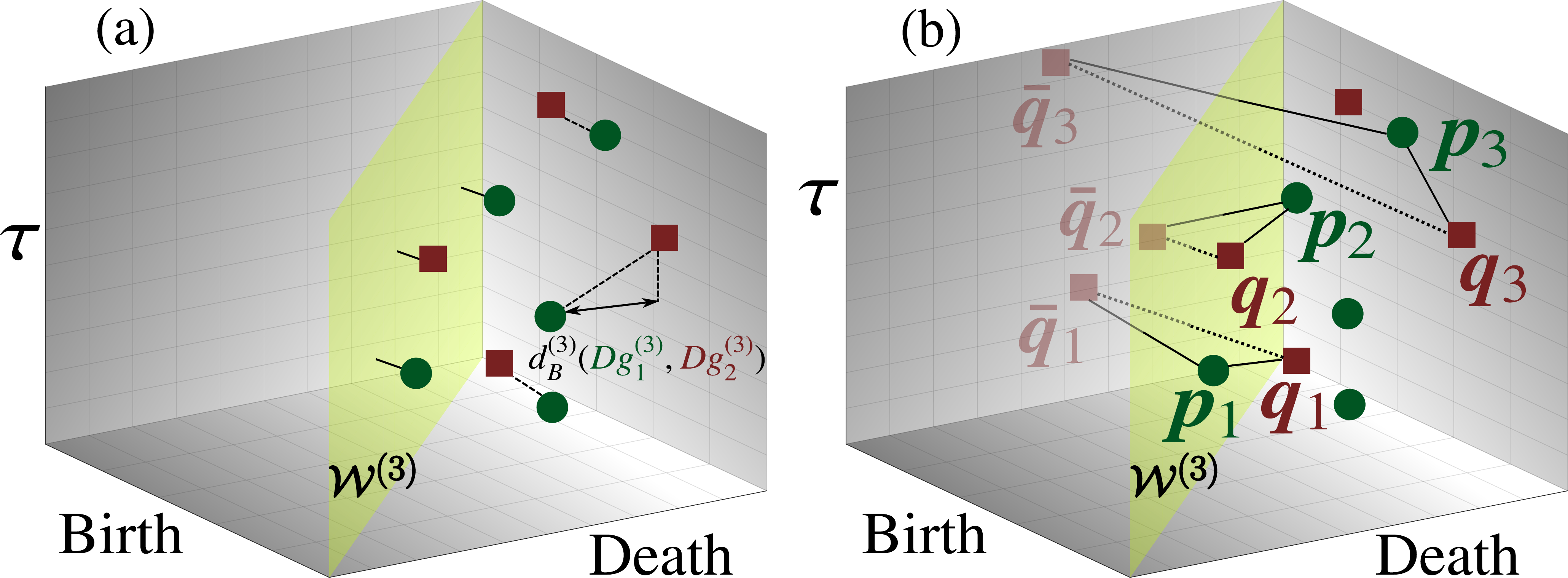}
	\protect\caption{(a) Bottleneck distance 
		between two diagrams $Dg_1^{(3)}$ (green points) and $Dg_2^{(3)}$ (red points). The $\tau$ axis is rescaled with the positive coefficient $\xi$ to adjust the scale difference between the point-wise distance and time. A green (red) point is matched to either a red (green) point or its projection on the diagonal plane, $\mathcal{W}^{(3)}$. 
		The distance between two matched points is the longest edge of their differences along any coordinate dimension.
		The bottleneck distance is defined as the longest distance between two matched points that is minimal compared to all other matchings.
		(b) Kernel between 
		$Dg_1^{(3)}$ (green points) and $Dg_2^{(3)}$ (dark red points). 
		The faint red points are symmetric to the dark red points with respect to $\mathcal{W}^{(3)}$. 
	    Points near $\mathcal{W}^{(3)}$, i.e., $\bp_1, \bq_2$, exhibit less influence, 
		whereas the pairs of points far away from $\mathcal{W}^{(3)}$, 
		i.e., $(\bp_2, \bq_1), (\bp_2, \bq_3), (\bp_3, \bq_1), (\bp_3, \bq_3)$,
		make essential contributions to the kernel.
		\label{fig:bottleneck}}
\end{figure}

Time-series data tend to be considerably noisy, 
which is a feature that can be considered to be either
a measurement artifact or an inherent part of the dynamics themselves.
Therefore, the persistence diagram should be stable 
with respect to the data being perturbed by noise.
To evaluate the stability of the persistence diagram, 
we introduce the concept of \textit{bottleneck distance} 
as a metric structure for comparing the persistence diagrams.
Given two three-dimensional persistence diagrams $Dg_1^{(3)}$ and $Dg_2^{(3)}$, consider all matchings $\psi$ such that a point on one diagram can be matched 
either to a point on the other diagram 
or to its projection on the diagonal plane $\mathcal{W}^{(3)} = \{(b, b, \tau) \mid b,\tau \in \mathbb{R}\}$ (Fig.~\ref{fig:bottleneck}(a)). 
For each pair $(\bp, \bq) \in \psi$ for which $\bp=\left(b_1,d_1,\tau_1\right)$ and $\bq=\left(b_2,d_2,\tau_2\right)$, we define the \textit{relative infinity-norm distance} between $\bp$ and $\bq$ as $d^{(\infty)}_{\xi}(\bp, \bq)=\max\left(|b_1-b_2|, |d_1-d_2|, \xi|\tau_1-\tau_2|\right)$,
where $\xi$ is a positive rescaling coefficient introduced to adjust the scale difference between the point-wise distance and time.
The bottleneck distance $d^{(3)}_{B, \xi}(Dg_{1}^{(3)},Dg_{2}^{(3)})$ 
can be defined as the infimum of the longest matched relative infinity-norm distance over all matchings $\psi$:
\begin{equation}\label{eqn:bottleneck:def}
d^{(3)}_{B, \xi}(Dg_{1}^{(3)},Dg_{2}^{(3)}) = \inf_{\psi} \max_{(\bp,\bq) \in \psi} d^{(\infty)}_{\xi}(\bp, \bq).
\end{equation}

We show that three-dimensional persistence diagrams are stable with respect to the bottleneck distance under perturbation applied to the time series.
Given two time series $x(t)$ and $y(t)$ of the same length, 
let $Dg_{(m)}^{(3)}(x)$ and $Dg_{(m)}^{(3)}(y)$ be their three-dimensional persistence diagrams 
respectively, as calculated for the embedding dimension $m$.
Based on the stability properties of two-dimensional persistence diagrams~\cite{chazal:stability:2014}, 
we can prove the following stability property of three-dimensional pesistence diagrams (see Appendix~\ref{sec:appx:stab}):
\begin{equation}\label{eqn:bottleneck:stablility}
d^{(3)}_{B, \xi}(Dg_{(m)}^{(3)}(x),Dg_{(m)}^{(3)}(y)) \leq 2\sqrt{m}\max_{t}|x(t)-y(t)|
\end{equation}
for an arbitrary positive $\xi$.
If we identify $y(t)$ as the perturbed data obtained by adding noise to $x(t)$,
Eq.~\eqref{eqn:bottleneck:stablility} shows that the upper limit of 
the bottleneck distance between $Dg_{(m)}^{(3)}(x)$ and $Dg_{(m)}^{(3)}(y)$ 
is governed by the magnitude of the noise.
Thus, the inequality of Eq.~\eqref{eqn:bottleneck:stablility} states that 
our three-dimensional persistence diagrams are robust with respect 
to the time-series data being perturbed by noise.
Therefore, these diagrams can be used as discriminating features for characterizing the time series.

\subsection{Kernel method for the topological features}
To use the persistence diagrams as features for performing the statistical learning tasks, such as classification, 
we must define a similarity measure, such as kernel mapping, 
for such diagrams~\cite{reininghaus:mskernel:2015, kusano:gskernel:2016, carrire:wskernel:2017}. 
Because the points that are close to $\mathcal{W}^{(3)}$ can be considered to be insignificant topological features,
they should not influence the computed value of this kernel. 
Given the positive bandwidth $\sigma$ and the positive rescaling parameter $\xi$, 
the kernel $k_{\sigma, \xi}$ between two three-dimensional persistence diagrams $Dg^{(3)}_1$ and $Dg^{(3)}_2$ can be defined as
\begin{align}\label{eqn:kernel:def}
&k_{\sigma, \xi}(Dg^{(3)}_1, Dg^{(3)}_2)\nonumber\\
&=\frac{1}{\sigma\sqrt{2\pi}}\sum_{\bp\in Dg^{(3)}_1, \bq \in Dg^{(3)}_2} \left( e^{-\frac{d_{\xi}^2(\bp, \bq)}{2\sigma^2}} - e^{-\frac{d_{\xi}^2(\bp, \bbq)}{2\sigma^2}} \right),
\end{align}
where $\bbq$ is a symmetric point of $\bq$ with respect to $\mathcal{W}^{(3)}$ and $d^2_{\xi}(\bp, \bq)=|b_1-b_2|^2+|d_1-d_2|^2+\xi^2|\tau_1-\tau_2|^2$,  $d^2_{\xi}(\bp, \bbq)=|b_1-d_2|^2+|d_1-b_2|^2+\xi^2|\tau_1-\tau_2|^2$, with $\bp=\left(b_1,d_1,\tau_1\right)$ and $\bq=\left(b_2,d_2,\tau_2\right)$.
If $\bp$ or $\bq$ is near $\mathcal{W}^{(3)}$, i.e., $b_1 \approx d_1$ or $b_2 \approx d_2$, then $d^2_{\xi}(\bp, \bq) \approx d^2_{\xi}(\bp, \bbq)$ and the pair $(\bp, \bq)$ will have a minor influence on
the value of the kernel (Fig.~\ref{fig:bottleneck}(b)).
Based on Ref.~\cite{reininghaus:mskernel:2015}, we can prove that $k_{\sigma,\xi}$ is a positive-definite kernel (see Appendix~\ref{sec:appx:kernel}).
When we apply $k_{\sigma, \xi}$ to time-series classification, 
the parameters $\sigma$ and $\xi$ are observed to affect the classification performance 
and can be chosen either by cross-validation or in a heuristic manner~\cite{gretton:nips:2007} (see Appendix~\ref{sec:appx:kernel}).
In our classification tasks, we use the normalized version of the kernel, which can be calculated as
$
k_{\sigma, \xi}(E, F) \leftarrow k_{\sigma, \xi}(E, F) / \sqrt{k_{\sigma, \xi}(E, E)k_{\sigma, \xi}(F, F)},
$
where $E$ and $F$ denote two persistence diagrams.
The source code that has been used to calculate the persistence diagrams and kernels can be found on GitHub~\cite{gitsource}.

\section{Results}
\subsection{Classification of the synthetic oscillatory and non-oscillatory data}
Initially, the proposed method is applied 
to classify periodic and aperiodic time series using synthetic single-cell data. 
This is challenging because of the difficulty associated with discriminating 
between an oscillation containing noise and a mere noisy fluctuation~\cite{phillips:hesplos:2017}.
We generate synthetic mRNA and protein time-series data from a stochastic model of the \textit{Hes1} genetic 
oscillator~\cite{monk:hesmodel:2003, galla:fluctuation:2009} exhibiting 
negative autoregulation with delay. We use the delayed version of the Gillespie  
algorithm~\cite{gillespie:exact:1977, anderson:gillespie:2007} 
to generate data from 1,000 cells in both the oscillatory and non-oscillatory parameter regimes~\cite{galla:fluctuation:2009, brett:distributed:2013}. 
We measure the protein levels after every $\nu$ (= 64, 32, 16, 8) min for 4,096 min. 
We normalize the time series to have zero mean and unit variance;
further, we add Gaussian white noise with variance $\sigma^2_n=0.1$ to assess the robustness of the method.

\begin{figure*}
	\includegraphics[width=15cm]{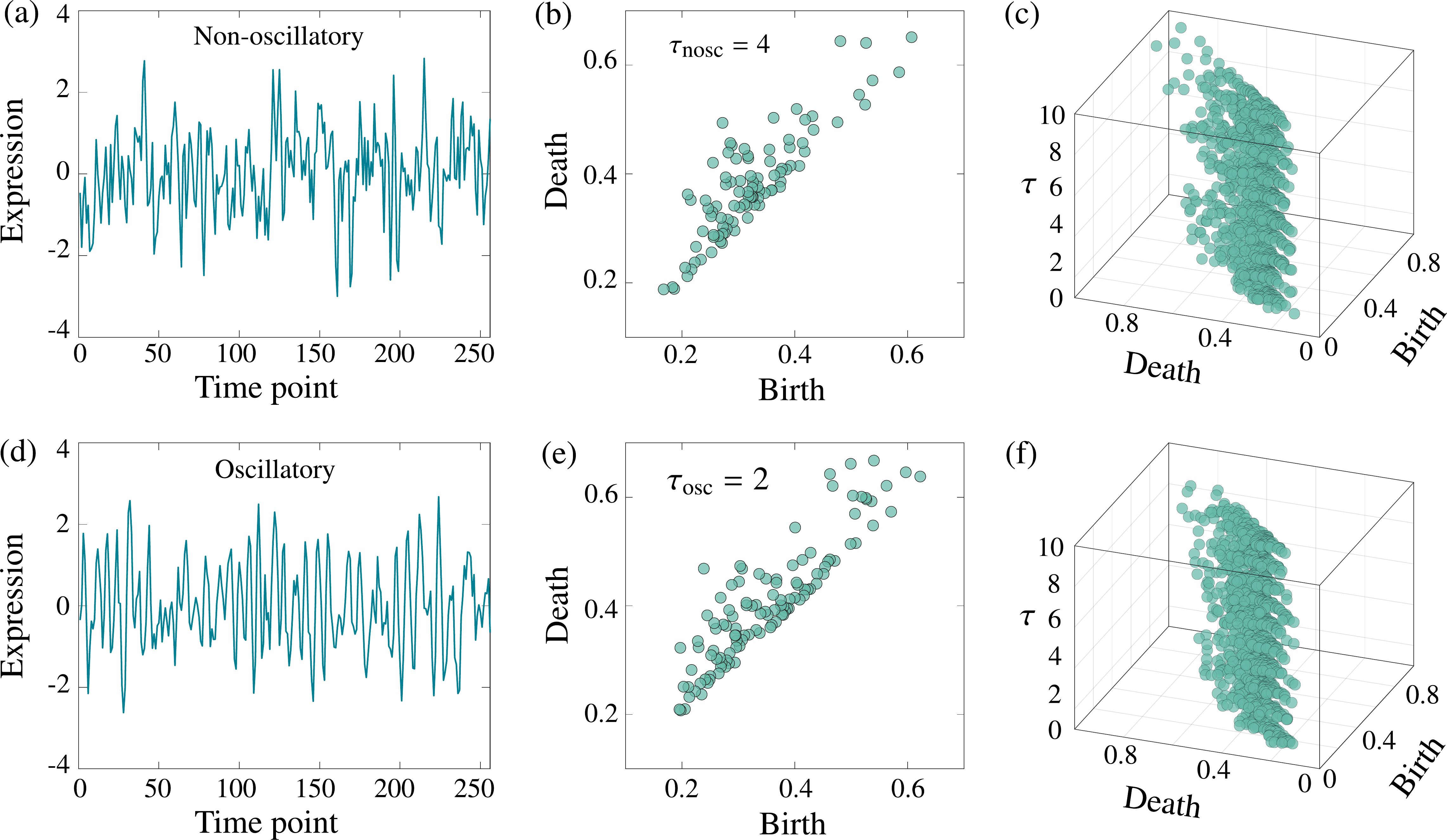}
	\protect\caption{(a)(d) Examples of two time series generated from the (a) non-oscillatory and (d) oscillatory regimes of the Hes1 model measured after every 16 min.
		(b)(e) Two-dimensional persistence diagrams of the loop patterns calculated with embedding dimension $m=3$ and a single value of $\tau$  
		selected according to~\cite{fraser:pra:1986}
        for the (b) non-oscillatory ($\tau=4$) and (e) oscillatory ($\tau=2$) time series.
		(c)(f) Three-dimensional persistence diagrams of loop patterns calculated with $m=3$ and $10$ values of $\tau$ given by $\tau_1=1, \tau_2=2, \ldots, \tau_{10}=10$ for the (c) non-oscillatory and (f) oscillatory time series.
		\label{fig:hesmodel_ts}}
\end{figure*}

Figure~\ref{fig:hesmodel_ts} depicts the examples of the time series and persistence diagrams 
for the two regimes with an embedding dimension of $m=3$. 
Figures~\ref{fig:hesmodel_ts}(a) and (d) show the time series generated by the measurements 
in the non-oscillatory and oscillatory regimes, respectively, after every $\nu=16$ min,
and Figs.~\ref{fig:hesmodel_ts}(b) and (e) denote their respective two-dimensional persistence diagrams 
computed from the single-delay embedding.
In Figs.~\ref{fig:hesmodel_ts}(b) and (e), $\tau$ 
is selected by the mutual-information method
to maximize the statistical measure of nonlinear independence 
in the delay coordinates of the embedded points~\cite{fraser:pra:1986}.
Figures~\ref{fig:hesmodel_ts}(c) and (f) depict the three-dimensional 
persistence diagrams obtained using Fig.~\ref{fig:hesmodel_ts}(a) and (d), respectively, 
with $K=10$ time-delay values as $\tau_1=1,\tau_2=2,\ldots,\tau_{10}=10$.
In these examples, it is difficult to distinguish between the two regimes 
using either the original time series or the two-dimensional diagrams, 
whereas there is a distinct pattern in the three-dimensional diagrams 
because all the topological variations are considered when $\tau$ changes. 
In the three-dimensional diagram of the non-oscillatory data, the points,
which represent the loops in the embedded space,
are extensively distributed along the birth and death scales, 
corresponding to the random fluctuation in the time series.
In contrast, in the oscillatory data,
points are densely distributed along the birth and death scales;
this manifests the appearance of repeated loop patterns in the embedded space,
thereby indicating periodic patterns in the time series.

\begin{figure}
	\includegraphics[width=8.5cm]{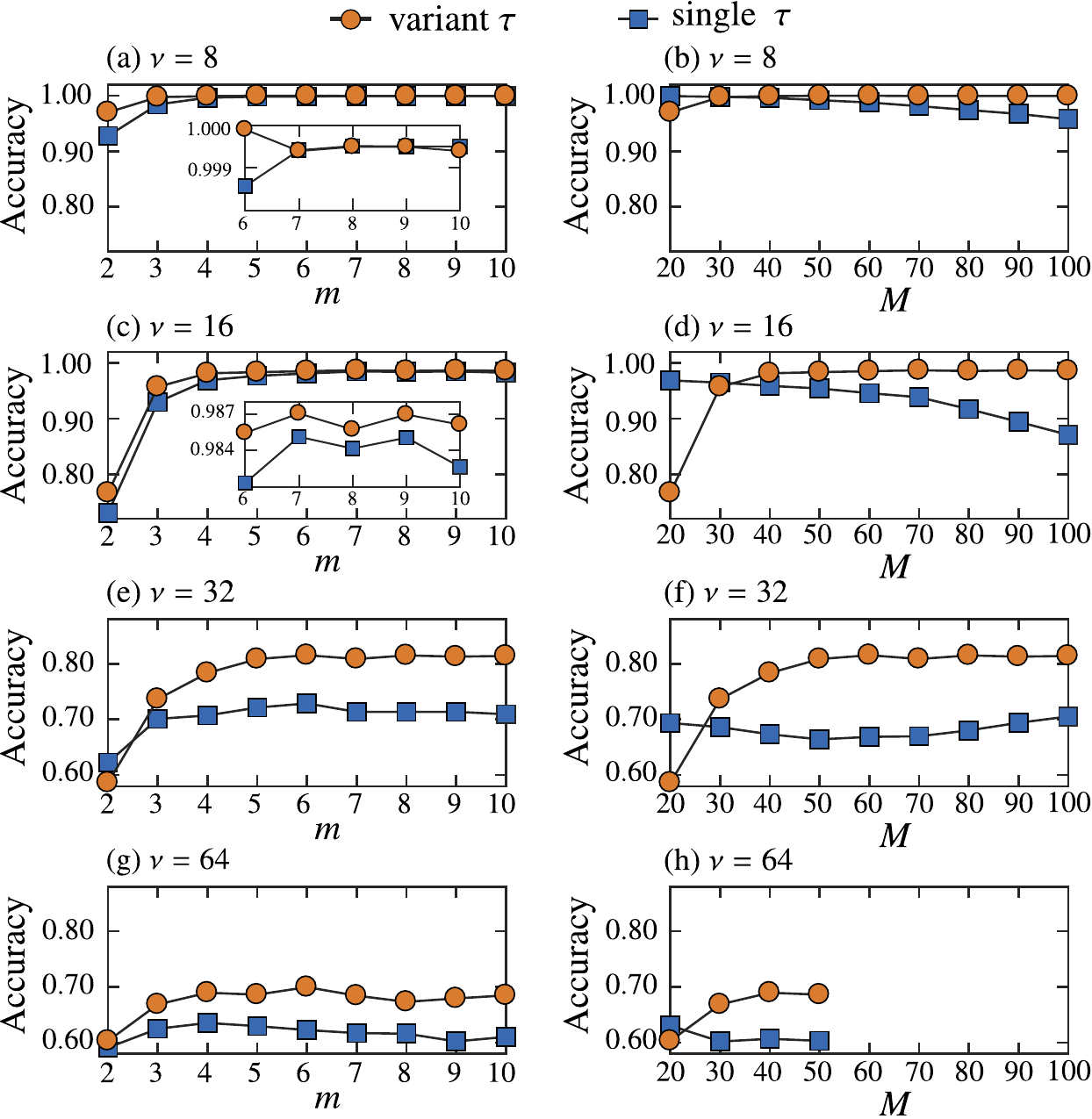}
	
	\protect\caption{
	Average classification accuracies of the delay-variant (circles) and single-delay (squares) 
	methods over 100 random splits of the Hes1-model data 
	at different (a)(c)(e)(g) embedding dimensions $m$ and different (b)(d)(f)(h) effective embedding dimensions $M$.
	The time-series data were obtained at different
	measurement intervals (a)(b) $\nu=8$, (c)(d) $\nu=16$, (e)(f) $\nu=32$, and (g)(h) $\nu=64$.
    The insets in (a) ($\nu=8$) and (c) ($\nu=16$) highlight the accuracy over the embedding dimension m=$6\text{--}10$.}
		\label{fig:hesmodel_acc}
\end{figure}

Further, we assess the effectiveness of the proposed topological features by classifying the data obtained from the Hes1 model, which can be randomly split with equal probability to belong to the training and test sets.
We use the support vector machine~\cite{prml:bishop:2006} to perform classification in the kernel space. 
In the delay-variant method, we consider $K=10$ time-delay values as $\tau_1=1,\tau_2=2,\ldots,\tau_{10}=10$.
In the single-delay method, we choose a single value of $\tau$ for each value from $\{\tau_1,\ldots,\tau_{10} \}$.
For each embedding dimension of the single-delay method, 
we use the maximum accuracy in the test dataset over 
all the values of $\tau$ for performing comparison with the delay-variant method.
Figure~\ref{fig:hesmodel_acc} depicts the average classification accuracy obtained using 100 random splits 
at different embedding dimensions $m$ and different measurement intervals 
$\nu=8$ (Fig.~\ref{fig:hesmodel_acc}(a)), $\nu=16$ (Fig.~\ref{fig:hesmodel_acc}(c)), 
$\nu=32$ (Fig.~\ref{fig:hesmodel_acc}(e)), and $\nu=64$ (Fig.~\ref{fig:hesmodel_acc}(g)).
When the same embedding dimension $m$ is being used, the total dimension for 
the delay-variant method ($mK$) is observed to be higher than that for the single-delay method ($m$). 
In the classification task, the high-dimensional features tend to achieve high classification accuracies. 
Therefore, to ensure that a fair comparison between the single-delay method and the delay-variant method can be performed, 
we define the effective embedding dimension as $M=mK$ 
and further compare the accuracy for different values of $M$ with $\nu=8$ (Fig.~\ref{fig:hesmodel_acc}(b)), $\nu=16$ (Fig.~\ref{fig:hesmodel_acc}(d)), 
$\nu=32$ (Fig.~\ref{fig:hesmodel_acc}(f)),
and $\nu=64$ (Fig.~\ref{fig:hesmodel_acc}(h)).
For $\nu=64$ (Fig.~\ref{fig:hesmodel_acc}(h)), 
we do not consider embeddings with $M \geq 60$ 
because the length of the time series is 64.

When the comparison is performed over the embedding dimension $m$, 
Fig.~\ref{fig:hesmodel_acc} depicts that 
both the methods will achieve almost the same accuracy
when the time series is long enough ($\nu=8, 16$) at $m \geq 4$.
In the single-delay method, 
$\tau$ should be selected by cross-validation for performing a fair comparison;
however, even with trial and error to obtain the maximum accuracy in the test dataset
in the single-delay method,
our delay-variant method still performs better.
When $\nu$ is increased, the time series is reduced, 
thereby increasing the difficulty of classification;
regardless, the delay-variant method attains a higher accuracy than the single-delay method.
For $\nu=32, 64$, the delay-variant method outperforms the single-delay method 
over a wide range of $m$ ($4\text{--}10$) in terms of the accuracy. 
This observation indicates that the single-delay method degrades quickly
with a shorter time series because it is highly sensitive to the choice of $\tau$.
However, by considering a range of $\tau$ with delay embedding,
this sensitivity can be reduced, thereby achieving stable performance.

While performing the comparison over the effective embedding dimension $M$, 
the accuracy of the delay-variant method does not change 
over a wide range of $M$ ($40\text{--}100$), 
indicating that this method is more reliable than the single-delay method 
(Fig.~\ref{fig:hesmodel_acc}(b), (d), (f), and (h)).
For a given effective dimension, the embedded points of 
single-delay method become highly folded 
and excessively sparse in a high-dimensional space
because the embedding dimension that is used in the single-delay method 
is considerably higher than that used in the delay-variant method.
Furthermore, selecting a considerably high embedding dimension 
increases the computational complexity of the embedding
and causes the high impact of noise acting in a high 
proportion of elements in delay coordinates.
Therefore, selecting a considerably high embedding dimension makes it
disadvantageous to use topological features for characterizing the time series,
thereby degrading the classification performance.

\begin{figure}
	\includegraphics[width=8.5cm]{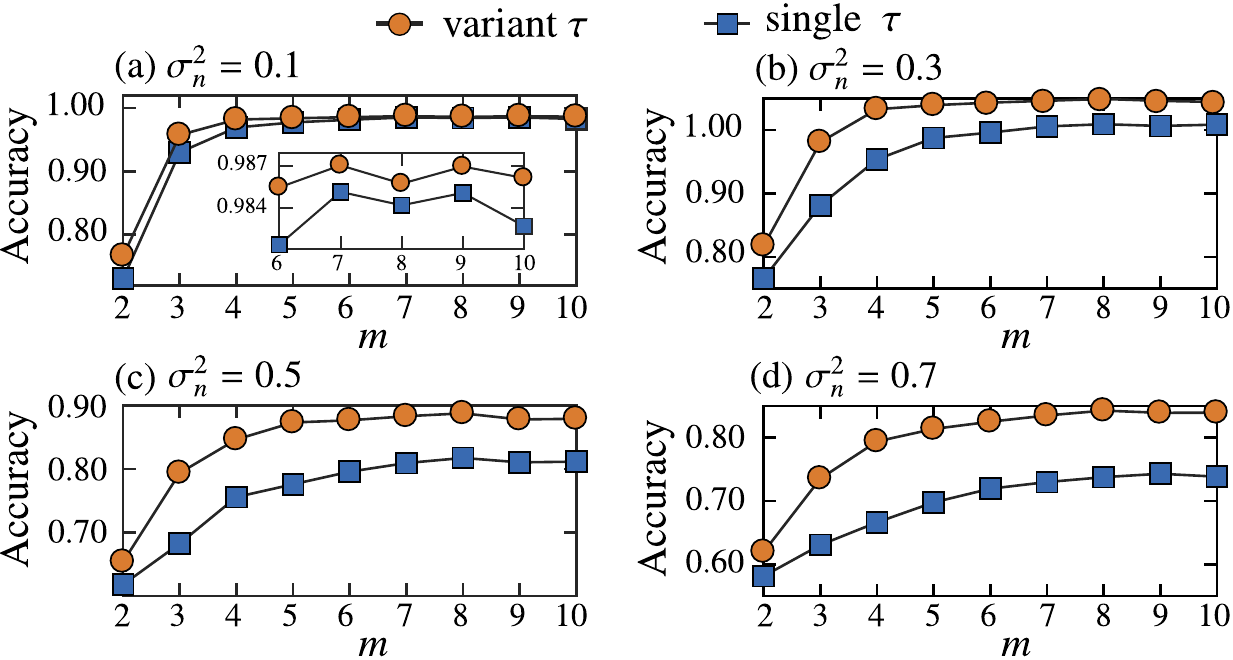}
	
	\protect\caption{
	Average classification accuracies of the delay-variant (circles) and single-delay (squares) methods 
	over 100 random splits of the Hes1-model data at different embedding dimensions $m$.
	The time-series data were obtained by adding the Gaussian white noise 
	with variance  (a) $\sigma^2_n=0.1$, (b) $\sigma^2_n=0.3$, (c) $\sigma^2_n=0.5$, and (d) $\sigma^2_n=0.7$
	to the time series generated with $\nu=16$ min.
	The inset in (a) ($\sigma^2_n=0.1$) depicts the accuracy over the embedding dimension $m=6\text{--}10$}
		\label{fig:hesmodel_acc:noise}
\end{figure}

To demonstrate that the delay-variant method is more robust than the single-delay method 
when the time series is being perturbed by noise,
we compare the two aforementioned methods based on the values of the noise variance $\sigma^2_n$ 
of 0.1, 0.3, 0.5, and 0.7.
Figure~\ref{fig:hesmodel_acc:noise} depicts the average classification accuracy 
versus the embedding dimension $m$ over 100 random splits 
at a measurement interval of $\nu=16$~min when the noise variance
$\sigma^2_n=0.1$ (Fig.~\ref{fig:hesmodel_acc:noise}(a)), 
$\sigma^2_n=0.3$ (Fig.~\ref{fig:hesmodel_acc:noise}(b)), 
$\sigma^2_n=0.5$ (Fig.~\ref{fig:hesmodel_acc:noise}(c)),
and $\sigma^2_n=0.7$ (Fig.~\ref{fig:hesmodel_acc:noise}(d)).
As $\sigma^2_n$ is increased, the classification becomes more difficult 
because the time-series data become noisier.
Even so, the delay-variant method still outperforms the single-delay method over $m$. 
Furthermore, the single-delay method degrades more quickly as $\sigma^2_n$
is increased, thereby indicating that our delay-variant method is more robust and stable
in noisy conditions than the single-delay method.

\subsection{Classification of the real time-series data}
Further, we evaluate the performance of the delay-variant method 
in classifying different heartbeat-signal patterns based on six real electrocardiogram (ECG) datasets in~\cite{UCRarchive}, 
namely ECG200 (separate normal and myocardial infarction heartbeats; 200 time series of length 96), 
ECG5000 (separate five levels of congestive heart failure; 5,000 time series of length 140), 
ECGFiveDays (separate records from two different days for the same patient; 884 time series of length 136), 
TwoLeadECG (separate records from two different leads; 1,162 time series of length 82), 
and Non-InvasiveFetalECGThorax1 and Non-InvasiveFetalECGThorax2 
(separate records from the left and right thorax with expert labeling 
of 42 classes of non-invasive fetal ECG; 3,765 time series of length 750 in each dataset).
We use the delay-variant method to classify the Caenorhabditis elegans roundworms from EigenWorms dataset
as either wild or mutant based on their movements~\cite{brown:elegans:791,yemini:database:2013}.
The movement trajectories are processed as 259 time series of length 900.
Finally, we denote that our method can be used to diagnose whether a certain symptom
can be observed in an automotive subsystem using the time-series data associated with the engine noise
with the FordB dataset for 4,446 time series of length 500~\cite{UCRarchive}.
Here, the training data in the FordB dataset were collected under typical operating conditions, 
whereas the test data were collected under noisy conditions.
For these datasets, we employ the train-test split provided in~\cite{UCRarchive}.
For the Non-InvasiveFetalECGThorax1, Non-InvasiveFetalECGThorax2, EigenWorms, and FordB datasets,
we downsampled the time series with sampling rates as 2, 2, 3, and 2, respectively,
before computing the persistence diagrams.

In the delay-variant method, we use $K=10$ time-delay values of $\tau_1=1,\tau_2=2,\ldots,\tau_{10}=10$
with embedding dimensions of $m=2,3,\ldots,10$.
In the single-delay method, we choose single $\tau$ for each value from $\{\tau_1,\ldots,\tau_{10} \}$.
For each embedding dimension of the single-delay method, 
we use the maximum accuracy in the test dataset over 
all values of $\tau$ to compare with the delay-variant method.
In both the methods, we use a linear combination of normalized kernels for the zero-dimensional and one-dimensional holes.
The combination weights and the embedding dimension $m$ are selected by cross-validation (see Appendix~\ref{sec:appx:kernel}).
Further, we compare the delay-variant and single-delay methods
along with some alternative standard approaches.
Most of the previous research on time-series classification 
focused on finding appropriate similarity measures for the $1^{\text{st}}$-nearest neighbor (NN) classifier. 
For ensuring similarity in the time domain, we consider either the Euclidean (E) or dynamic time warping (D) distance.
For ensuring similarity in the frequency domain and for similarity in the autocorrelation, 
we consider the power spectrum (PS) and the autocorrelation function (AC). 
We also examine the elastic ensemble (EE)~\cite{lines:ee:2014} 
as a combination of nearest neighbor classifiers using multiple distance measures in the time domain.
Finally, we perform the learned-shapelets (LS) method, 
which classifies time series by learning the representative \textit{shapelets} (i.e., short discriminant time-series subsequences)~\cite{grabocka:LST:2014}.
Instead of implementing these standard algorithms, 
we use the results from Ref.~\cite{bagnall:review:2016,tscresults}.
The test results are presented in Table~\ref{tab:acc}, 
wherein the best and second-best accuracy scores of each dataset 
are colored in dark pink and light pink, respectively. 
The delay-variant method shows better results than
the single-delay method for all the datasets and outperforms all the other algorithms on an average. 
Further, the delay-variant method offers the best results for four of the eight datasets 
and the second-best results for the remaining four,
suggesting that our method is an effective mechanism to classify the time series.
\begin{table}
	\caption{\label{tab:acc} Classification accuracies (\%) for the ECG200, ECG5000, 
	Non-InvasiveFetalECGThorax1 (Thorax1), Non-InvasiveFetalECGThorax2 (Thorax2),
	ECGFiveDays (FiveDays), TwoLeadECG (TwoLead), 
	EigenWorms (Worms), and FordB datasets. 
		For each dataset, the best and second-best scores are colored in dark pink and light pink, respectively. 
		The notations are  $1^{\text{st}}$-nearest neighbor classifier (NN), 
		Euclidean distance (E), dynamic time warping distance (D), autocorrelation function (AC), 
		power spectrum (PS), elastic ensemble (EE), and learned-shapelets method (LS).}
	\begin{ruledtabular}
		\begin{tabular}{lcccccccc}
			\st{Data\\~}     &\textbf{\st{Delay\\variant}} &\st{Single\\delay} &\st{NN\\(E)} &\st{NN\\(D)} &\st{NN\\(AC)}  &\st{NN\\(PS)}        &EE       &LS      \\ \hline
			ECG200   &\rst 90.0                  &87.0               &\rnd 88.0          &\rnd 88.0          &82.0 &86.0  &\rnd 88.0  &\rnd 88.0\\
            ECG5000  &\rnd 93.6                  &92.1               &92.5          &92.5          &91.0 &\rnd 93.6  &\rst 93.9  &93.2\\
			Thorax1  &\rst 91.8                  &78.2               &82.9          &82.9          &72.1 &\rnd 87.5  &84.6       &25.9\\
			Thorax2  &\rst 93.0                  &83.6               &88.0          &87.0          &75.2 &88.4       &\rnd 91.4  &77.0\\
            FiveDays &\rnd 99.9                  &92.0               &79.7          &79.7          &98.1 &\rst 100.0 &82.0       &\rst 100.0\\
			TwoLead  &\rnd 99.4                  &94.0               &74.7          &86.8          &80.4 &96.1       &97.1       &\rst 99.6\\
			Worms &\rst 83.1                &\rst 83.1               &61.0          &58.4          &76.6 &\rnd 81.8  &68.8       &72.7\\
			FordB      &\rnd 90.8                &78.2                  &60.6          &59.9          &78.0 &79.0  &66.2       &\rst 91.7\\
		\end{tabular}
	\end{ruledtabular}
\end{table}

\subsection{Multiple-time-scale patterns captured by delay-variant embedding}
Subsequently, we investigate the types of features that can be effectively captured using the delay-variant method. 
Because the time scale of the extracted pattern is partially dependent on the time delay,
the delay-variant method should be able to extract the patterns that comprise multiple different time scales.
To better understand this ability, we analyze the synthetic noisy time series obtained from a frequency-modulated model. 
Given the original signal $s_m(t)=A_m\sin(2\pi f_mt)$ with a carrier signal $s_c(t)=A_c\sin(2\pi f_c t)$,
we consider the modulated signal to be $s(t)=A_c\sin(2\pi f_c t+A_m\sin(2\pi f_mt))$, 
where $t\in[0, 0.1]$, $A_m=10$, $A_c=1.0$, and $f_m=25$. 
Further, we consider the frequency $f_c$ to be $f_c=5f$ with $f=1\text{--}20$ to
represent patterns with multiple time scales in the modulated signal. 
For each value of $f_c$, we generate 20 noisy discrete time series $z_l(n)=s(0.0002n)+w_l(n)$, 
where $l=1,2,\ldots,20$, $n=0,1,\ldots,500$, and $w_l(n)$ denote the Gaussian white noise with a variance of 0.1.
\begin{figure}
	\includegraphics[width=8.5cm]{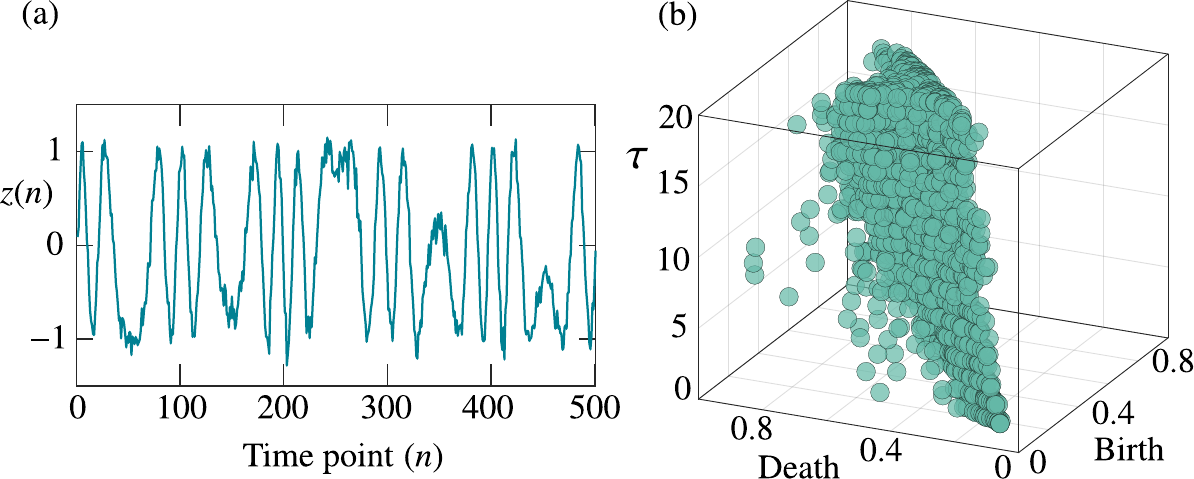}
	\protect\caption{(a) Time series generated from the frequency-modulated model. (b) Three-dimensional persistence diagram of the loop patterns calculated with $m=6$ and 20 values of $\tau$ given by $\tau_1=1, \tau_2=2, \ldots, \tau_{20}=20$. The multiple-time-scale components in the time series corresponding to the limit-cycle oscillations of the dynamics are revealed from the points in the diagram over different values of $\tau$.
		\label{fig:FM:eg}}
\end{figure}

In the delay-variant method, we use $K=20$ time-delay values of $\tau_1=1, \tau_2=2,\ldots,\tau_{20}=20$.
Figure~\ref{fig:FM:eg} depicts a time series with $f_c=25$ and 
its three-dimensional persistence diagram for the loops or cycles at an embedding dimension of $m=6$.
In the persistence diagram, the points near the diagonal plane 
$\mathcal{W}^{(3)} = \{(b, b, \tau) \mid b,\tau \in \mathbb{R}\}$ 
represent the noise in the time series,
whereas the points that are away from $\mathcal{W}^{(3)}$ 
represent significant cycles in the embedded space.
Therefore, the points in the diagram that has been plotted for different values of $\tau$ 
reveal the multiple-time-scale components in the time series corresponding to the limit-cycle oscillations of the dynamics.

\begin{figure*}
	\includegraphics[width=16cm]{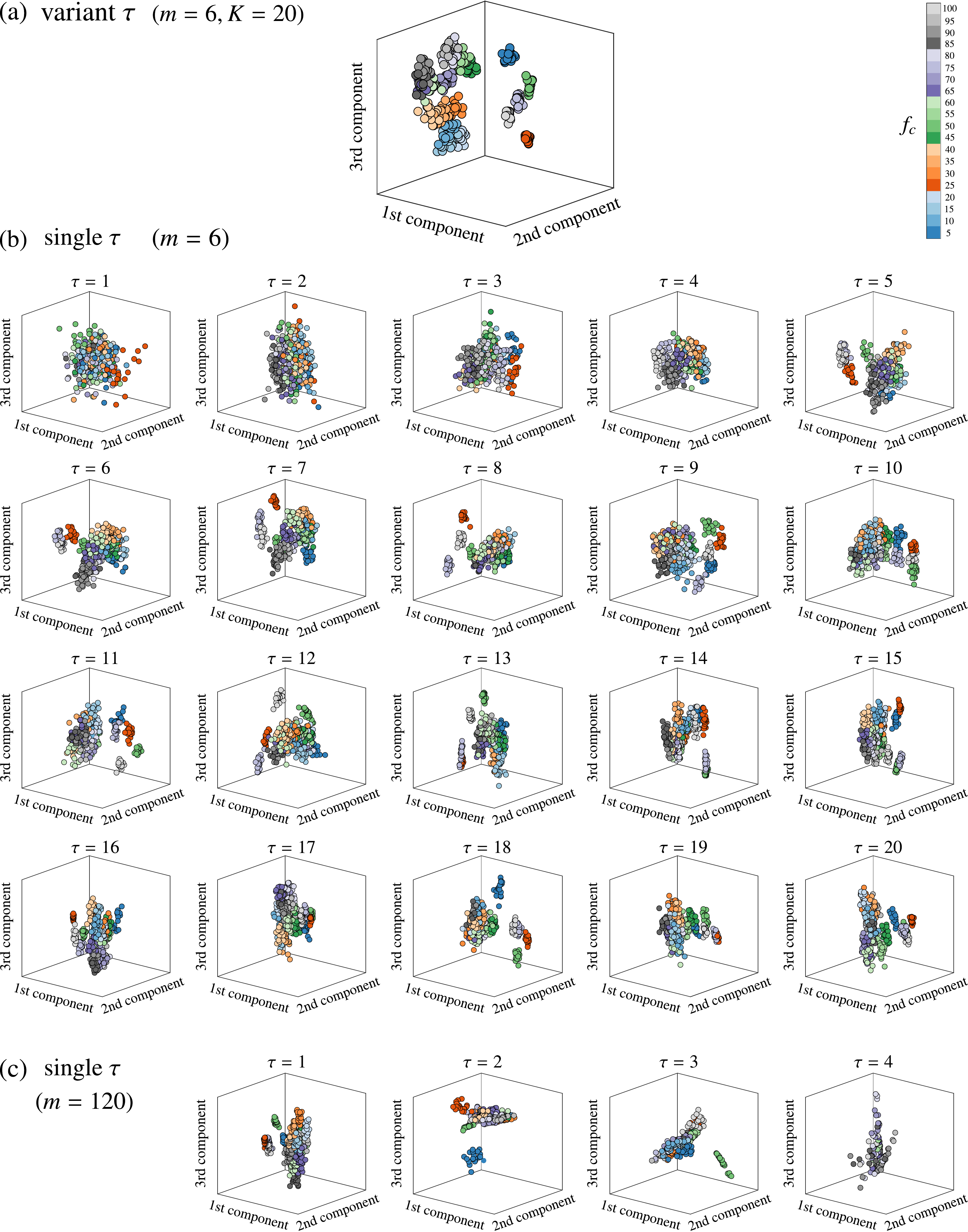}
	
	\protect\caption{Projection of the features of synthetic frequency-modulated time series 
	from kernel space to principal components for 
	    (a) the delay-variant method with embedding dimension $m=6$,
        (b) the single-delay method with $m=6$ and $\tau=1,2,\ldots,20$,
        and (c) the single-delay method with $m=120$
        (for performing comparison with the delay-variant method at the same effective embedding dimension $M=120$)
        and $\tau=1,2,3,4$.
		Different colors represent the data generated using different values of the carrier frequency $f_c$.}
        \label{fig:FM_kpca}
\end{figure*}

An example of the principal component projection from the kernel space is depicted in Fig.~\ref{fig:FM_kpca} for 
(a) the delay-variant method with embedding dimension $m=6$,
(b) the single-delay method with $m=6$ and $\tau=1,2,\ldots,20$,
and (c) the single-delay method with $m=120$
(to compare with the delay-variant method at the same effective embedding dimension $M=120$).
Because the length of the time series is 501, in the single-delay approach, 
$\tau$ can only take values from $\{1, 2, 3, 4\}$ when $m=120$.
Different colors represent the data generated using different values of $f_c$.
Figure~\ref{fig:FM_kpca} depicts that the delay-variant method results in more distinct regions 
corresponding to different values of $f_c$ than that obtained using the single-delay method.
Note that the points obtained using the single-delay method cannot be distinct, regardless of the angle.
These results denote that the delay-variant method is superior to the single-delay method 
with respect to the ability to identify changes in the multiple-time-scale patterns in time series.
\section{Concluding remarks}
We have demonstrated that the topological features that are constructed using delay-variant embedding 
can capture the topological variation in a time series when the time-delay value changes. 
Therefore, these features can be used to discriminate between different time series.
Theoretically, we have mathematically denoted that these features are robust 
when the time series is being perturbed by noise.
Our method outperformed the standard time-series analysis techniques while classifying both synthetic and real time-series data.
These results indicate that the topological features that are deduced with
delay-variant embedding can be used to reveal the representative features of the original time series.

In general, the mathematical model of a time series denotes a stochastic process.
It is important to understand the factor that distinguishes this model
from other models with respect to the dynamics.
Recent powerful deep learning methods 
have focused on accurately classifying time series 
with predefined labels without identifying the essential behavior of the model.
Unfortunately, these approaches could fail 
when the input is slightly but deliberately perturbed,
resulting in misclassification by a neural network.
Such perturbed examples can be referred to as \textit{adversarial examples}
and have gained considerable attention recently in the discussion about the robustness of deep learning~\cite{szegedy:adversarial:2014}.
Machine learning tools, such as deep learning methods, 
find it difficult to understand why adversarial examples are misclassified
because of the black-box nature of the causality between the input data and their underlying model.
From another viewpoint, topological data analysis provides a systematic methodology 
for understanding the true behavior of the data, 
which will eventually become significant
in characterizing the behavior of the model.

We expect that our study will lead to a unified analysis mechanism of the time-series data.
This study paves several opportunities for the application of topological tools 
to produce effective algorithms for analyzing the time-series data. 
For instance, given a time series, we could use delay-variant embedding
to predict its future values or to detect anomalous values or outliers therein.

\appendix
\section{Topological features obtained from data\label{sec:topo:features}}
	
\begin{figure*}[t]
	\includegraphics[width=16.0cm]{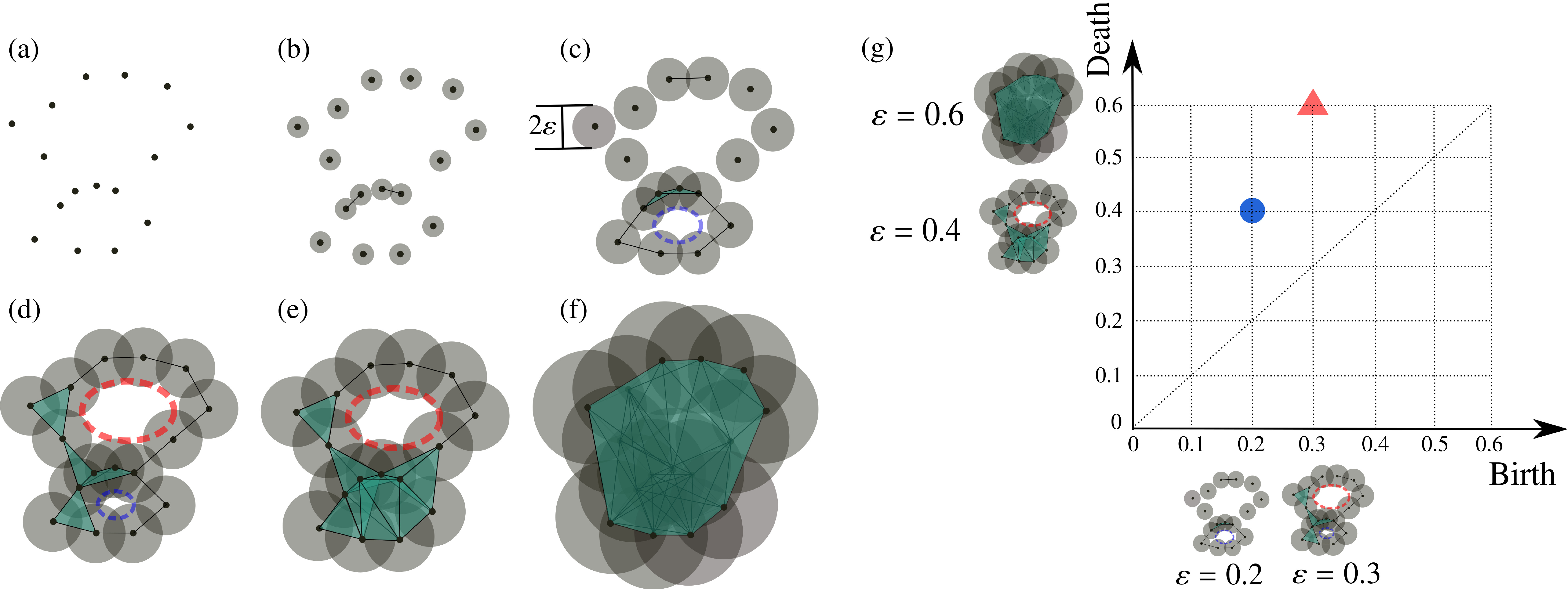}
	\protect\caption{We consider $L$-dimensional hyperspheres with radius $\varepsilon$ centered at each point. From (a) to (f), we increase the radius $\varepsilon$ gradually. By increasing $\varepsilon$, holes appear and disappear in the region. The first one-dimensional hole (blue loop) appears at (c), while the second (red loop) appears at (d). Then at (e), the first hole disappears and finally the second one disappears at (f).
	(g) An examplary two-dimensional persistence diagram calculated by considering the appearance and disappearance of loops from a filtration of the Vietoris--Rips complex. The radius $\varepsilon$ takes the discrete values from the set \{0, 0.1, 0.2, 0.3, 0.4, 0.5, 0.6\}. There are two persistence pairs (0.2, 0.4) and (0.3, 0.6), which represent the appearance and disappearance of the blue loop and the red loop, respectively. 
	These persistence pairs are displayed as a blue circle and a red triangle in the Cartesian plane, respectively. The collection of all the persistence pairs in the filtration is a two-dimensional persistence diagram.
	\label{fig:pointcloud}}
\end{figure*}

    To extract the topological features from a set $P$ of points in the Euclideand space $\mathbb{R}^L$, we build a $\varepsilon$-scale Vietoris--Rips complex (denoted as $\text{VR}(P, \varepsilon)$) from a union of $L$-dimensional hyperspheres of radius $\varepsilon$ centered at each point in $P$. Every collection of $n+1$ affinely independent points in $P$ forms an $n$-simplex in $\text{VR}(P, \varepsilon)$ if the pairwise distance between points is less than or equal to $2\varepsilon$. The complex $\text{VR}(P, \varepsilon)$ gives us the topological information from $P$ associated with radius $\varepsilon$. For example, in Fig.~\ref{fig:pointcloud}(d), there are two loops (called one-dimensional holes) and one connected component (called a zero-dimensional hole). However, this information depends on how to choose the radius $\varepsilon$. If $\varepsilon$ is too small, the complex created by union hyperspheres (Fig.~\ref{fig:pointcloud}(b)) remains almost the same as the discrete points (Fig.~\ref{fig:pointcloud}(a)). If $\varepsilon$ is too large, we obtain a trivially connected and overlapped complex without any hole inside it (Fig. \ref{fig:pointcloud}(f)).

	The problem above can be solved by considering not only a single radius, but all choices of radius $\varepsilon$.
    This yields a \emph{filtration}, 
    which is a sequence of simplicial complexes used to monitor the appearance of holes such as clusters and loops over changing $\varepsilon$. For example, in Fig.~\ref{fig:pointcloud}, we start from $\varepsilon=0$ (Fig.~\ref{fig:pointcloud}(a)); then we increase $\varepsilon$ gradually to see whether the holes appear or disappear. 
    When a hole appears and disappears at radii $\varepsilon=b$ and $\varepsilon=d$, respectively, the hole is characterized by a pair $(b, d)$, where $b$, $d$, and $(b, d)$ are referred to as the \textit{birth scale}, \textit{death scale}, and \textit{persistence pair}, respectively. 
    
    The persistence pairs, which we use as topological features, 
    are displayed in the Cartesian plane as a two-dimensional persistence diagram where the birth and death scales appear as the horizontal and vertical coordinates, respectively. In the two-dimensional persistence diagram, points far from the diagonal generally correspond to robust features, which are persistent over a long time, whereas those near the diagonal are regarded as noise in the data. We provide an exemplary two-dimensional persistence diagram in Fig.~\ref{fig:pointcloud}(g), where we consider the appearance and disappearance of loops in a filtration of the Vietoris--Rips complex constructed from points when $\varepsilon$ takes the discrete values from the set $\{0, 0.1, 0.2, 0.3, 0.4, 0.5, 0.6 \}$.

\section{Proof of the stability of the three-dimensional persistence diagrams\label{sec:appx:stab}}
\renewcommand{\theproposition}{A\arabic{proposition}}
		We prove the result in Eq.~\eqref{eqn:bottleneck:stablility}. 
		First, we define the bottleneck distance $d^{(2)}_B$ between two two-dimensional persistence diagrams $Dg_1^{(2)}$ and $Dg_2^{(2)}$ as
		\begin{equation}\label{eqn:bottleneck2:def}
	d^{(2)}_B(Dg_{1}^{(2)},Dg_{2}^{(2)}) = \inf_{\gamma} \max_{(\bp,\bq) \in \gamma} \Vert \bp - \bq\Vert_{\infty},
	\end{equation} 
	where $\gamma$ is a matching between $Dg_1^{(2)}$ and $Dg_2^{(2)}$
    such that a point on a diagram is matched to a point on the other or to its projection on the diagonal line $\mathcal{W}^{(2)} = \{(b, b) \mid b \in \mathbb{R}\}$. 
    Here the distance $\Vert \boldsymbol{z} \Vert_{\infty}$ is defined as $\Vert \boldsymbol{z} \Vert_{\infty}=\max\{|z_1|,\ldots, |z_n|\}$ for $\boldsymbol{z} = (z_1,\ldots, z_n) \in \mathbb{R}^n$.
	
    The bottleneck distance between the two-dimensional persistence diagrams satisfies the following property~\cite{chazal:stability:2014}: 
	
	\begin{proposition}
		Let $X$ and $Y$ be finite sets of points embedded in the Euclidean space $\mathbb{R}^n$. Denote their two-dimensional persistent diagrams as $Dg^{(2)}(X)$ and $Dg^{(2)}(Y)$, respectively. Then,
		\begin{equation}
		d^{(2)}_B(Dg^{(2)}(X), Dg^{(2)}(Y)) \leq 2d_H(X, Y),
		\end{equation}
		where $d_H(X, Y)$ is the Hausdorff distance given by 
		\begin{equation}
		 d_H(X, Y) = \max \left\{ \max_{\bx\in X} \min_{\by \in Y} d(\bx, \by),  \max_{\by \in Y} \min_{\bx \in X}d(\bx, \by)\right\}\nonumber.
		\end{equation}
		Here, $d(\bx, \by)$ is the Euclidean distance between $\bx, \by$.
	\end{proposition}
	
	For each $\tau \in \mathcal{T}=\{\tau_1,\tau_2,...,\tau_K\}$, we denote $\xmt$ and $\ymt$ are the embedded points of $x(t)$ and $y(t)$, respectively, in an embedding space with dimension $m$ and time delay $\tau$. 
	Consider two two-dimensional persistence diagrams $Dg^{(2)}(\xmt)$ and $Dg^{(2)}(\ymt)$ calculated from $\xmt$ and $\ymt$, respectively. 
	Let $\Gamma_{\tau}$ be the set of matchings defined in Eq.~\eqref{eqn:bottleneck2:def} between $Dg^{(2)}(\xmt)$ and $Dg^{(2)}(\ymt)$. 
	For each collection $\Lambda=\{\gamma_1, \gamma_2, \ldots, \gamma_K  \mid \gamma_i \in \Gamma_{\tau_i}, i=1,2,\ldots, K\}$, we construct a matching $\psi$ between two three-dimensional persistence diagrams $\dggm(x)$ and $\dggm(y)$, such that, 
	for each $(\bp, \bq) \in \psi$, then $\bp=(b_1, d_1, \tau)$, $\bq=(b_2, d_2, \tau)$,
	and $(\bp_{\gamma}, \bq_{\gamma}) \in \gamma$, where $\bp_{\gamma}=(b_1, d_1)$, $\bq_{\gamma}=(b_2, d_2)$, and $\gamma \in \Lambda \cap \Gamma_{\tau}$.
	Let $\Gamma$ be the set of all matchings $\psi$ constructed this way. 
	From the definition of bottleneck distance, we have the following inequality:
	 
	\begin{equation} \label{eqn:bttdis:inf}
		d^{(3)}_{B, \xi}(\dggm(x),\dggm(y)) \leq \inf_{\psi \in \Gamma} \max_{(\bp, \bq) \in \psi} d^{(\infty)}_{\xi}(\bp, \bq).
	\end{equation}
    
	For $(\bp, \bq) \in \psi$, we have
	\begin{align}
			d^{(\infty)}_{\xi}(\bp, \bq) &= \max\{|b_1-b_2|, |d_1-d_2|, \xi|\tau-\tau|\}\\
			 &= \max\{|b_1-b_2|, |d_1-d_2|\}\\
			 &= \Vert \bp_{\gamma} - \bq_{\gamma}\Vert_{\infty},
	\end{align}
	
    and Eq.~\eqref{eqn:bttdis:inf} becomes
	\begin{align} 
		d^{(3)}_{B, \xi}&(\dggm(x),\dggm(y)) \nonumber \\
		& \leq \max_{\tau \in \mathcal{T}} \inf_{\gamma \in \Gamma_{\tau}} \max_{(\bp_{\gamma}, \bq_{\gamma}) \in \gamma} \Vert \bp_{\gamma} - \bq_{\gamma}\Vert_{\infty}\\
	 &= \max_{\tau \in \mathcal{T}} d^{(2)}_B(Dg^{(2)}(\xmt), Dg^{(2)}(\ymt)).\label{eqn:bttdis:2to3}
	\end{align}
	
	From Eq.~\eqref{eqn:bttdis:2to3} and Proposition A1, we have
	\begin{equation} \label{eqn:stab:bottle}
	d^{(3)}_{B, \xi}(\dggm(x),\dggm(y)) \leq 2 \max_{\tau \in \mathcal{T}} d_H(\xmt, \ymt).
	\end{equation}
	At each time point $t_0$, we consider $\mathcal{X}_{(m)}^{\tau}(t_0)=\left( x(t_0), x(t_0-\tau), \ldots, x(t_0-(m-1)\tau)\right) \in \xmt$ 
		and $\mathcal{Y}_{(m)}^{\tau}(t_0)=\left( y(t_0), y(t_0-\tau), \ldots, y(t_0-(m-1)\tau) \right) \in \ymt$.
	From the definition of Euclidean distance, we have
	\begin{align} 
	d(\mathcal{X}_{(m)}^{\tau}(t_0), \mathcal{Y}_{(m)}^{\tau}(t_0)) &= \sqrt{\sum_{i=0}^{m-1}[x(t_0+i\tau)-y(t_0+i\tau)]^2} \nonumber\\
	&\leq \sqrt{m}\max_{t}| x(t)-y(t) | \label{eqn:euc:diff}.
	\end{align}
	
	From Eq.~\eqref{eqn:euc:diff} and the definition of Hausdorff distance, we have the following inequality:\begin{widetext}
	\begin{align} 
		d_H(X_{(m)}^\tau, Y_{(m)}^{\tau})\nonumber &= \max \left\{ \max_{t_1} \min_{t_2} d(\mathcal{X}_{(m)}^{\tau}(t_1), \mathcal{Y}_{(m)}^{\tau}(t_2)),  \max_{t_2} \min_{t_1} d(\mathcal{X}_{(m)}^{\tau}(t_1), \mathcal{Y}_{(m)}^{\tau}(t_2))\right\} \\
		 &\leq \max \left\{ \max_{t_1}d(\mathcal{X}_{(m)}^{\tau}(t_1), \mathcal{Y}_{(m)}^{\tau}(t_1)),  \max_{t_2}d(\mathcal{X}_{(m)}^{\tau}(t_2), \mathcal{Y}_{(m)}^{\tau}(t_2))\right\}\leq \sqrt{m}\max_{t} | x(t)-y(t) |.\label{eqn:stab:bottle2}
	\end{align}
	\end{widetext}
	Equation~\eqref{eqn:bottleneck:stablility} is obtained by taking the maximum of Eq.~\eqref{eqn:stab:bottle2} over $\tau \in \mathcal{T}$ and applying it to Eq.~\eqref{eqn:stab:bottle}.

\section{Kernel of the three-dimensional persistence diagrams\label{sec:appx:kernel}}
\subsection{Proof of the positive-definite property}
	We prove that the proposed kernel for the three-dimensional persistence diagrams is positive-definite.
	We prove for the case when the positive rescaling coefficient $\xi=1$. The proof is straightforward for other positive values of $\xi$.
	
	For each parameter $\sigma$ and a three-dimensional persistence diagram $Dg^{(3)}$, we define the following feature mapping $\Phi_{\sigma}: \mathcal{D}^{(3)}\rightarrow L^2(\Omega)$, where $\mathcal{D}^{(3)}$ is the space of three-dimensional persistence diagrams, $\Omega=\{\bx=(x_1, x_2, x_3) \mid x_1, x_2, x_3 \in \mathbb{R}, x_1 \leq x_2\}$ and $L^2(\Omega)$ is the Hilbert space of square-integrable $L^2$-functions defined on the domain $\Omega$: 
	\begin{align}
	 &\Phi_{\sigma}(Dg^{(3)})(\bx) \nonumber\\
	 &= \frac{1}{\sqrt{\kappa}}\sum_{\bp \in Dg^{(3)}}\exp\left(-\frac{\|\bx-\bp\|^2}{\sigma^2}\right) - \exp\left(-\frac{\|\bx-\bbp\|^2}{\sigma^2}\right),
	\end{align}
	where $\bbp$ is a symmetric point of $\bp$ with respect to diagonal plane $\mathcal{W}^{(3)}$ on $\mathbb{R}^3$ and $\kappa$ is a positive value depending on $\sigma$, which we will show later.
	
	We show that the kernel defined in Eq.~\eqref{eqn:kernel:def} is the inner product of $\Phi_{\sigma}$ on $L^2(\Omega)$ as
	\begin{align}
	k_{\sigma,\xi}(Dg^{(3)}_1, Dg^{(3)}_2) &= \langle \Phi_{\sigma}(Dg^{(3)}_1), \Phi_{\sigma}(Dg^{(3)}_2) \rangle_{L^2(\Omega)} \label{eqn:in:1} \\
	&= \int_{\Omega}\Phi_{\sigma}(Dg^{(3)}_1)(\bx)\Phi_{\sigma}(Dg^{(3)}_2)(\bx)d\bx \label{eqn:in:2}.
	\end{align}
	
	We extend the domain of function  $\Phi_{\sigma}(Dg^{(3)})(\bx)$ from $\Omega$ to $\mathbb{R}^3$ to obtain a function that is symmetric with respect to the diagonal plane $\mathcal{W}^{(3)}$ (because $\Vert\bx - \bp \Vert = \Vert \bbx - \bbp \Vert$ and $\Vert \bx - \bbp \Vert = \Vert \bbx - \bp \Vert$).
	Then we have
	\begin{align}
	\int_{\Omega^{(3)}}&\Phi_{\sigma}(Dg^{(3)}_1)(\bx)\Phi_{\sigma}(Dg^{(3)}_2)(\bx)d\bx \nonumber\\
	&=\frac{1}{2}\int_{\mathbb{R}^3}\Phi_{\sigma}(Dg^{(3)}_1)(\bx)\Phi_{\sigma}(Dg^{(3)}_2)(\bx)d\bx\nonumber\\
	&= \frac{1}{2\kappa}\sum_{\substack{\bp\in Dg^{(3)}_1\\ \bq \in Dg^{(3)}_2}}\left[A(\bp, \bq) + A(\bbp, \bbq) - A(\bbp, \bq) - A(\bp, \bbq)\right], \label{eqn:ker:mid}
	\end{align}
	where $A(\bc, \bd)$ is defined as
	\begin{align}
	A(\bc, \bd)=&\int_{\mathbb{R}^3}\exp\left(-\frac{\|\bx-\bc\|^2+\|\bx-\bd\|^2}{\sigma^2}\right)d\bx\nonumber\\
	=&\frac{\sigma^3(2\pi)^{3/2}}{8}\exp\left(-\frac{\|\bc-\bd\|^2}{2\sigma^2}\right).\label{eqn:int:mid}
	\end{align}
	Here, $\bc, \bd \in \mathbb{R}^3$. Since $\|\bp-\bq\|^2=\|\bbp-\bbq\|^2, \|\bbp-\bq\|^2=\|\bp-\bbq\|^2$, and from Eqs.~\eqref{eqn:ker:mid} and~\eqref{eqn:int:mid}
	we have the closed form of the kernel where $\kappa=\dfrac{\pi^2\sigma^4}{2}$.
	
	The kernel is positive-definite due to the inner-product nature of the feature mapping. Consider the three-dimensional persistence diagrams $Dg^{(3)}_1, \ldots, Dg^{(3)}_N$ of the $l$-dimensional holes that are needed to compute the kernel. The Gram matrix for these diagrams is defined as $\mathbf{K}^{(l)}_{N\times N}=[k_{ij}]$, whose element $k_{ij}$ is $k_{ij}=k_{\sigma,\xi}(Dg^{(3)}_i, Dg^{(3)}_j)$ with $i=1,\ldots,N$ and $j=1,\ldots,N$.
	
\subsection{Selection of the kernel parameters}
	In time-series classification experiments, we compute the kernel by taking time-delay values as discrete values with sampling interval 1 and set the rescaling coefficient to $\xi=\sigma$. 
	The selection of kernel bandwidth $\sigma$ can be chosen by cross-validation; however, as proposed in~\cite{gretton:nips:2007}, we present here a heuristic way to select $\sigma$.
	Consider the three-dimensional persistence diagrams $Dg_1^{(3)}, Dg_2^{(3)}, \ldots, Dg_N^{(3)}$ that are required to compute the kernel. We denote $\sigma^2_s = \text{median}\{(b_i-b_j)^2+(d_i-d_j)^2 \mid (b_i, d_i, \tau_i), (b_j, d_j, \tau_j) \in Dg_s^{(3)}\}$ with $s=1,2,\ldots,N$. 
	$\sigma$ is set as $\sigma^2 = \dfrac{1}{2}\text{median}\{\sigma^2_s \mid s = 1,\ldots,N\}$, such that $2\sigma^2$ takes values close to many $(b_i-b_j)^2+(d_i-d_j)^2$ values.

\subsection{Using kernels with holes of multiple dimensions}
	For all $l$-dimensional holes, we obtain the persistence diagrams and compute their kernels as $k_{\sigma_l, \xi_l}$, where $\sigma_l$ is the bandwidth of this kernel and $\xi_l$ is the positive rescaling coefficient corresponding with $l$-dimensional holes. To use persistence diagrams for different dimensions of holes, we can combine the kernels at various dimensions through linear combinations. In our time-series classification experiments, we only consider the topological features of zero-dimensional holes (connected components) and one-dimensional holes (loops). Thus, the combined Gram matrix of $N$ data can be defined as
	\begin{equation}
	\mathbf{K}_{N\times N}=\alpha_0\mathbf{K}^{(0)}_{N\times N} + \alpha_1\mathbf{K}^{(1)}_{N\times N},
	\end{equation}
	where $0 \leq \alpha_0, \alpha_1 \leq 1, \alpha_0 + \alpha_1=1$, and $\mathbf{K}^{(0)}_{N\times N}$ and $\mathbf{K}^{(1)}_{N\times N}$ are the Gram matrices of $N$ persistence diagrams at the zero-dimensional and one-dimensional holes, respectively. In our time-series classification experiments, we choose $\alpha_0$ from 0, 0.0001, 0.0002, 0.0005, 0.001, 0.002, 0.005, 0.01, 0.02, 0.05, 0.1, 0.2, 0.5, 1.0 by cross-validation.
	
\section{Synthetic data\label{sec:appx:synthetic}}
	We generate data from Hes1 regulatory model, which is a stochastic model of the \emph{Hes1} genetic oscillator exhibiting negative autoregulation with delay~\cite{monk:hesmodel:2003,galla:fluctuation:2009}. The model describes the concentrations and interactions of two types of particles: \textit{hes}1 mRNA molecules, denoted by M, and Hes1 protein molecules, denoted by P. The stochastic dynamics are defined by the following reactions: 
	\begin{eqnarray}
	M&\stackrel{\mu_m}{\longrightarrow}&\varnothing; \label{eq:r1}\\
	P&\stackrel{\mu_p}{\longrightarrow}&\varnothing; \label{eq:r2}\\
	M&\stackrel{\beta_p}{\longrightarrow}&M+P; \label{eq:r3} \\
	\varnothing&\stackrel{g(n_p), K(\zeta)}{\Longrightarrow}&M.\label{eq:r4}
	\end{eqnarray}	
	
	The degradations of mRNA and protein are described in reactions~\eqref{eq:r1} and~\eqref{eq:r2}, respectively, where the rates of these degradation are $\mu_m $ and $\mu_p$, respectively. mRNA molecules are translated into protein via reaction~\eqref{eq:r3} by the translation-rate parameter $\beta_p$. The final reaction,~\eqref{eq:r4} with a double arrow describes the transcription process for producing mRNA which is accompanied by time delay. The rate of \textit{hes}1 mRNA production depends on the concentration of Hes1 protein molecules through a negative-feedback mechanism, as described by the function $g(n_p) = \beta_m \left[1+\left[n_p / (P_0\Theta) \right]^h\right]^{-1}$.
	Here, $n_p$ is the number of protein molecules in the system, $\beta_m$, $P_0$, and $h$ are constants, and $\Theta$ is the size of the system.
	This transcription process is associated with a time delay $\zeta$ drawn from a distribution $K(\zeta)$; that is, the protein concentration at time $t-\zeta$ only affects the production of mRNA at time $t$. 
	
	In the simulations, the protein levels were measured after every $\nu$ (= 64, 32, 16, 8) min for 4,096 min. 
	The measurements start at $t=5,000$ min for the system to equilibrate.
	The parameters for the oscillatory regime are $P_0=300, h=1, \zeta=0, \beta_m=\beta_p=1, \mu_m=\mu_p=0.07, \Theta=20$, and for the non-oscillatory regime are $P_0=100, h=3, \zeta=18, \beta_m=\beta_p=1, \mu_m=\mu_p=0.03, \Theta=20$.


\begin{thebibliography}{41}%
	\makeatletter
	\providecommand \@ifxundefined [1]{%
		\@ifx{#1\undefined}
	}%
	\providecommand \@ifnum [1]{%
		\ifnum #1\expandafter \@firstoftwo
		\else \expandafter \@secondoftwo
		\fi
	}%
	\providecommand \@ifx [1]{%
		\ifx #1\expandafter \@firstoftwo
		\else \expandafter \@secondoftwo
		\fi
	}%
	\providecommand \natexlab [1]{#1}%
	\providecommand \enquote  [1]{``#1''}%
	\providecommand \bibnamefont  [1]{#1}%
	\providecommand \bibfnamefont [1]{#1}%
	\providecommand \citenamefont [1]{#1}%
	\providecommand \href@noop [0]{\@secondoftwo}%
	\providecommand \href [0]{\begingroup \@sanitize@url \@href}%
	\providecommand \@href[1]{\@@startlink{#1}\@@href}%
	\providecommand \@@href[1]{\endgroup#1\@@endlink}%
	\providecommand \@sanitize@url [0]{\catcode `\\12\catcode `\$12\catcode
		`\&12\catcode `\#12\catcode `\^12\catcode `\_12\catcode `\%12\relax}%
	\providecommand \@@startlink[1]{}%
	\providecommand \@@endlink[0]{}%
	\providecommand \url  [0]{\begingroup\@sanitize@url \@url }%
	\providecommand \@url [1]{\endgroup\@href {#1}{\urlprefix }}%
	\providecommand \urlprefix  [0]{URL }%
	\providecommand \Eprint [0]{\href }%
	\providecommand \doibase [0]{http://dx.doi.org/}%
	\providecommand \selectlanguage [0]{\@gobble}%
	\providecommand \bibinfo  [0]{\@secondoftwo}%
	\providecommand \bibfield  [0]{\@secondoftwo}%
	\providecommand \translation [1]{[#1]}%
	\providecommand \BibitemOpen [0]{}%
	\providecommand \bibitemStop [0]{}%
	\providecommand \bibitemNoStop [0]{.\EOS\space}%
	\providecommand \EOS [0]{\spacefactor3000\relax}%
	\providecommand \BibitemShut  [1]{\csname bibitem#1\endcsname}%
	\let\auto@bib@innerbib\@empty
	\bibitem [{\citenamefont {Donato}\ \emph {et~al.}(2016)\citenamefont {Donato},
		\citenamefont {Gori}, \citenamefont {Pettini}, \citenamefont {Petri},
		\citenamefont {De~Nigris}, \citenamefont {Franzosi},\ and\ \citenamefont
		{Vaccarino}}]{pre:donato:phase:16}%
	\BibitemOpen
	\bibfield  {author} {\bibinfo {author} {\bibfnamefont {I.}~\bibnamefont
			{Donato}}, \bibinfo {author} {\bibfnamefont {M.}~\bibnamefont {Gori}},
		\bibinfo {author} {\bibfnamefont {M.}~\bibnamefont {Pettini}}, \bibinfo
		{author} {\bibfnamefont {G.}~\bibnamefont {Petri}}, \bibinfo {author}
		{\bibfnamefont {S.}~\bibnamefont {De~Nigris}}, \bibinfo {author}
		{\bibfnamefont {R.}~\bibnamefont {Franzosi}}, \ and\ \bibinfo {author}
		{\bibfnamefont {F.}~\bibnamefont {Vaccarino}},\ }\href {\doibase
		10.1103/PhysRevE.93.052138} {\bibfield  {journal} {\bibinfo  {journal} {Phys.
				Rev. E}\ }\textbf {\bibinfo {volume} {93}},\ \bibinfo {pages} {052138}
		(\bibinfo {year} {2016})}\BibitemShut {NoStop}%
	\bibitem [{\citenamefont {Kusano}\ \emph {et~al.}(2016)\citenamefont {Kusano},
		\citenamefont {Fukumizu},\ and\ \citenamefont
		{Hiraoka}}]{kusano:gskernel:2016}%
	\BibitemOpen
	\bibfield  {author} {\bibinfo {author} {\bibfnamefont {G.}~\bibnamefont
			{Kusano}}, \bibinfo {author} {\bibfnamefont {K.}~\bibnamefont {Fukumizu}}, \
		and\ \bibinfo {author} {\bibfnamefont {Y.}~\bibnamefont {Hiraoka}},\ }in\
	\href {http://dl.acm.org/citation.cfm?id=3045390.3045602} {\emph {\bibinfo
			{booktitle} {Proc. 33th Int. Conf. Machine Learning (ICML)}}},\ Vol.~\bibinfo
	{volume} {48}\ (\bibinfo {year} {2016})\BibitemShut {NoStop}%
	\bibitem [{\citenamefont {Ardanza-Trevijano}\ \emph {et~al.}(2014)\citenamefont
		{Ardanza-Trevijano}, \citenamefont {Zuriguel}, \citenamefont {Ar\'evalo},\
		and\ \citenamefont {Maza}}]{pre:ardanza:granular:14}%
	\BibitemOpen
	\bibfield  {author} {\bibinfo {author} {\bibfnamefont {S.}~\bibnamefont
			{Ardanza-Trevijano}}, \bibinfo {author} {\bibfnamefont {I.}~\bibnamefont
			{Zuriguel}}, \bibinfo {author} {\bibfnamefont {R.}~\bibnamefont {Ar\'evalo}},
		\ and\ \bibinfo {author} {\bibfnamefont {D.}~\bibnamefont {Maza}},\ }\href
	{\doibase 10.1103/PhysRevE.89.052212} {\bibfield  {journal} {\bibinfo
			{journal} {Phys. Rev. E}\ }\textbf {\bibinfo {volume} {89}},\ \bibinfo
		{pages} {052212} (\bibinfo {year} {2014})}\BibitemShut {NoStop}%
	\bibitem [{\citenamefont {Nakamura}\ \emph {et~al.}(2015)\citenamefont
		{Nakamura}, \citenamefont {Hiraoka}, \citenamefont {Hirata}, \citenamefont
		{Escolar},\ and\ \citenamefont {Nishiura}}]{nakamura:nano:2015}%
	\BibitemOpen
	\bibfield  {author} {\bibinfo {author} {\bibfnamefont {T.}~\bibnamefont
			{Nakamura}}, \bibinfo {author} {\bibfnamefont {Y.}~\bibnamefont {Hiraoka}},
		\bibinfo {author} {\bibfnamefont {A.}~\bibnamefont {Hirata}}, \bibinfo
		{author} {\bibfnamefont {E.~G.}\ \bibnamefont {Escolar}}, \ and\ \bibinfo
		{author} {\bibfnamefont {Y.}~\bibnamefont {Nishiura}},\ }\href
	{http://stacks.iop.org/0957-4484/26/i=30/a=304001} {\bibfield  {journal}
		{\bibinfo  {journal} {Nanotechnology}\ }\textbf {\bibinfo {volume} {26}},\
		\bibinfo {pages} {304001} (\bibinfo {year} {2015})}\BibitemShut {NoStop}%
	\bibitem [{\citenamefont {Hiraoka}\ \emph {et~al.}(2016)\citenamefont
		{Hiraoka}, \citenamefont {Nakamura}, \citenamefont {Hirata}, \citenamefont
		{Escolar}, \citenamefont {Matsue},\ and\ \citenamefont
		{Nishiura}}]{hiraoka:pnas:2016}%
	\BibitemOpen
	\bibfield  {author} {\bibinfo {author} {\bibfnamefont {Y.}~\bibnamefont
			{Hiraoka}}, \bibinfo {author} {\bibfnamefont {T.}~\bibnamefont {Nakamura}},
		\bibinfo {author} {\bibfnamefont {A.}~\bibnamefont {Hirata}}, \bibinfo
		{author} {\bibfnamefont {E.~G.}\ \bibnamefont {Escolar}}, \bibinfo {author}
		{\bibfnamefont {K.}~\bibnamefont {Matsue}}, \ and\ \bibinfo {author}
		{\bibfnamefont {Y.}~\bibnamefont {Nishiura}},\ }\href
	{http://www.pnas.org/content/early/2016/06/07/1520877113} {\bibfield
		{journal} {\bibinfo  {journal} {Proc. Natl. Acad. Sci. U.S.A.}\ } (\bibinfo
		{year} {2016})}\BibitemShut {NoStop}%
	\bibitem [{\citenamefont {Ichinomiya}\ \emph {et~al.}(2017)\citenamefont
		{Ichinomiya}, \citenamefont {Obayashi},\ and\ \citenamefont
		{Hiraoka}}]{pre:ichinomiya:craze:17}%
	\BibitemOpen
	\bibfield  {author} {\bibinfo {author} {\bibfnamefont {T.}~\bibnamefont
			{Ichinomiya}}, \bibinfo {author} {\bibfnamefont {I.}~\bibnamefont
			{Obayashi}}, \ and\ \bibinfo {author} {\bibfnamefont {Y.}~\bibnamefont
			{Hiraoka}},\ }\href {\doibase 10.1103/PhysRevE.95.012504} {\bibfield
		{journal} {\bibinfo  {journal} {Phys. Rev. E}\ }\textbf {\bibinfo {volume}
			{95}},\ \bibinfo {pages} {012504} (\bibinfo {year} {2017})}\BibitemShut
	{NoStop}%
	\bibitem [{\citenamefont {Maleti{\'c}}\ \emph {et~al.}(2016)\citenamefont
		{Maleti{\'c}}, \citenamefont {Zhao},\ and\ \citenamefont
		{Rajkovi{\'c}}}]{maletic:dynamic:2016}%
	\BibitemOpen
	\bibfield  {author} {\bibinfo {author} {\bibfnamefont {S.}~\bibnamefont
			{Maleti{\'c}}}, \bibinfo {author} {\bibfnamefont {Y.}~\bibnamefont {Zhao}}, \
		and\ \bibinfo {author} {\bibfnamefont {M.}~\bibnamefont {Rajkovi{\'c}}},\
	}\href {\doibase 10.1063/1.4949472} {\bibfield  {journal} {\bibinfo
			{journal} {Chaos}\ }\textbf {\bibinfo {volume} {26}},\ \bibinfo {pages}
		{053105} (\bibinfo {year} {2016})}\BibitemShut {NoStop}%
	\bibitem [{\citenamefont {Mittal}\ and\ \citenamefont
		{Gupta}(2017)}]{mittal:bifucation:2017}%
	\BibitemOpen
	\bibfield  {author} {\bibinfo {author} {\bibfnamefont {K.}~\bibnamefont
			{Mittal}}\ and\ \bibinfo {author} {\bibfnamefont {S.}~\bibnamefont {Gupta}},\
	}\href {\doibase 10.1063/1.4983840} {\bibfield  {journal} {\bibinfo
			{journal} {Chaos}\ }\textbf {\bibinfo {volume} {27}},\ \bibinfo {pages}
		{051102} (\bibinfo {year} {2017})}\BibitemShut {NoStop}%
	\bibitem [{\citenamefont {Takens}(1981)}]{taken:1981}%
	\BibitemOpen
	\bibfield  {author} {\bibinfo {author} {\bibfnamefont {F.}~\bibnamefont
			{Takens}},\ }in\ \href {\doibase 10.1007/bfb0091924} {\emph {\bibinfo
			{booktitle} {Dynamical Systems and Turbulence}}},\ \bibinfo {series} {Lecture
		Notes in Mathematics}, Vol.\ \bibinfo {volume} {898}\ (\bibinfo  {publisher}
	{Springer-Verlag, Berlin},\ \bibinfo {year} {1981})\ pp.\ \bibinfo {pages}
	{366--381}\BibitemShut {NoStop}%
	\bibitem [{\citenamefont {Casdagli}\ \emph {et~al.}(1991)\citenamefont
		{Casdagli}, \citenamefont {Eubank}, \citenamefont {Farmer},\ and\
		\citenamefont {Gibson}}]{casdagli:state:1991}%
	\BibitemOpen
	\bibfield  {author} {\bibinfo {author} {\bibfnamefont {M.}~\bibnamefont
			{Casdagli}}, \bibinfo {author} {\bibfnamefont {S.}~\bibnamefont {Eubank}},
		\bibinfo {author} {\bibfnamefont {J.~D.}\ \bibnamefont {Farmer}}, \ and\
		\bibinfo {author} {\bibfnamefont {J.}~\bibnamefont {Gibson}},\ }\href
	{\doibase 10.1016/0167-2789(91)90222-U} {\bibfield  {journal} {\bibinfo
			{journal} {Physica D}\ }\textbf {\bibinfo {volume} {51}},\ \bibinfo {pages}
		{52} (\bibinfo {year} {1991})}\BibitemShut {NoStop}%
	\bibitem [{\citenamefont {Fraser}\ and\ \citenamefont
		{Swinney}(1986)}]{fraser:pra:1986}%
	\BibitemOpen
	\bibfield  {author} {\bibinfo {author} {\bibfnamefont {A.~M.}\ \bibnamefont
			{Fraser}}\ and\ \bibinfo {author} {\bibfnamefont {H.~L.}\ \bibnamefont
			{Swinney}},\ }\href {\doibase 10.1103/PhysRevA.33.1134} {\bibfield  {journal}
		{\bibinfo  {journal} {Phys. Rev. A}\ }\textbf {\bibinfo {volume} {33}},\
		\bibinfo {pages} {1134} (\bibinfo {year} {1986})}\BibitemShut {NoStop}%
	\bibitem [{\citenamefont {Liebert}\ and\ \citenamefont
		{Schuster}(1989)}]{liebert:proper:1989}%
	\BibitemOpen
	\bibfield  {author} {\bibinfo {author} {\bibfnamefont {W.}~\bibnamefont
			{Liebert}}\ and\ \bibinfo {author} {\bibfnamefont {H.}~\bibnamefont
			{Schuster}},\ }\href {\doibase 10.1016/0375-9601(89)90169-2} {\bibfield
		{journal} {\bibinfo  {journal} {Phys. Lett. A}\ }\textbf {\bibinfo {volume}
			{142}},\ \bibinfo {pages} {107} (\bibinfo {year} {1989})}\BibitemShut
	{NoStop}%
	\bibitem [{\citenamefont {Buzug}\ and\ \citenamefont
		{Pfister}(1992{\natexlab{a}})}]{buzug:optimal:1992}%
	\BibitemOpen
	\bibfield  {author} {\bibinfo {author} {\bibfnamefont {T.}~\bibnamefont
			{Buzug}}\ and\ \bibinfo {author} {\bibfnamefont {G.}~\bibnamefont
			{Pfister}},\ }\href {\doibase 10.1103/PhysRevA.45.7073} {\bibfield  {journal}
		{\bibinfo  {journal} {Phys. Rev. A}\ }\textbf {\bibinfo {volume} {45}},\
		\bibinfo {pages} {7073} (\bibinfo {year} {1992}{\natexlab{a}})}\BibitemShut
	{NoStop}%
	\bibitem [{\citenamefont {Buzug}\ and\ \citenamefont
		{Pfister}(1992{\natexlab{b}})}]{buzug:comp:1992}%
	\BibitemOpen
	\bibfield  {author} {\bibinfo {author} {\bibfnamefont {T.}~\bibnamefont
			{Buzug}}\ and\ \bibinfo {author} {\bibfnamefont {G.}~\bibnamefont
			{Pfister}},\ }\href {\doibase 10.1016/0167-2789(92)90104-U} {\bibfield
		{journal} {\bibinfo  {journal} {Physica D}\ }\textbf {\bibinfo {volume}
			{58}},\ \bibinfo {pages} {127 } (\bibinfo {year}
		{1992}{\natexlab{b}})}\BibitemShut {NoStop}%
	\bibitem [{\citenamefont {Rosenstein}\ \emph {et~al.}(1994)\citenamefont
		{Rosenstein}, \citenamefont {Collins},\ and\ \citenamefont
		{De~Luca}}]{rosenstein:recons:1994}%
	\BibitemOpen
	\bibfield  {author} {\bibinfo {author} {\bibfnamefont {M.~T.}\ \bibnamefont
			{Rosenstein}}, \bibinfo {author} {\bibfnamefont {J.~J.}\ \bibnamefont
			{Collins}}, \ and\ \bibinfo {author} {\bibfnamefont {C.~J.}\ \bibnamefont
			{De~Luca}},\ }\href {\doibase 10.1016/0167-2789(94)90226-7} {\bibfield
		{journal} {\bibinfo  {journal} {Physica D}\ }\textbf {\bibinfo {volume}
			{73}},\ \bibinfo {pages} {82} (\bibinfo {year} {1994})}\BibitemShut {NoStop}%
	\bibitem [{\citenamefont {Kantz}\ and\ \citenamefont
		{Schreiber}(2003)}]{kantz:nonlinear:2003}%
	\BibitemOpen
	\bibfield  {author} {\bibinfo {author} {\bibfnamefont {H.}~\bibnamefont
			{Kantz}}\ and\ \bibinfo {author} {\bibfnamefont {T.}~\bibnamefont
			{Schreiber}},\ }\href {\doibase 10.1017/CBO9780511755798} {\emph {\bibinfo
			{title} {Nonlinear Time Series Analysis}}},\ \bibinfo {edition} {2nd}\ ed.\
	(\bibinfo  {publisher} {Cambridge Univ. Press},\ \bibinfo {year}
	{2003})\BibitemShut {NoStop}%
	\bibitem [{\citenamefont {Bradley}\ and\ \citenamefont
		{Kantz}(2015)}]{bradley:nonlinear:2015}%
	\BibitemOpen
	\bibfield  {author} {\bibinfo {author} {\bibfnamefont {E.}~\bibnamefont
			{Bradley}}\ and\ \bibinfo {author} {\bibfnamefont {H.}~\bibnamefont
			{Kantz}},\ }\href {\doibase 10.1063/1.4917289} {\bibfield  {journal}
		{\bibinfo  {journal} {Chaos}\ }\textbf {\bibinfo {volume} {25}},\ \bibinfo
		{pages} {097610} (\bibinfo {year} {2015})}\BibitemShut {NoStop}%
	\bibitem [{\citenamefont {Carlsson}(2009)}]{carlsson:topology:2009}%
	\BibitemOpen
	\bibfield  {author} {\bibinfo {author} {\bibfnamefont {G.}~\bibnamefont
			{Carlsson}},\ }\href {\doibase 10.1090/S0273-0979-09-01249-X} {\bibfield
		{journal} {\bibinfo  {journal} {Bull. Amer. Math. Soc.}\ }\textbf {\bibinfo
			{volume} {46}},\ \bibinfo {pages} {255} (\bibinfo {year} {2009})}\BibitemShut
	{NoStop}%
	\bibitem [{\citenamefont {Edelsbrunner}\ and\ \citenamefont
		{Harer}(2010)}]{edels:topobook:2010}%
	\BibitemOpen
	\bibfield  {author} {\bibinfo {author} {\bibfnamefont {H.}~\bibnamefont
			{Edelsbrunner}}\ and\ \bibinfo {author} {\bibfnamefont {J.}~\bibnamefont
			{Harer}},\ }\href {\doibase 10.1090/mbk/069} {\emph {\bibinfo {title}
			{Computational Topology. An Introduction.}}}\ (\bibinfo  {publisher} {Amer.
		Math. Soc.},\ \bibinfo {year} {2010})\BibitemShut {NoStop}%
	\bibitem [{\citenamefont {Edelsbrunner}\ \emph {et~al.}(2002)\citenamefont
		{Edelsbrunner}, \citenamefont {Letscher},\ and\ \citenamefont
		{Zomorodian}}]{edelsbrunner:2002}%
	\BibitemOpen
	\bibfield  {author} {\bibinfo {author} {\bibfnamefont {H.}~\bibnamefont
			{Edelsbrunner}}, \bibinfo {author} {\bibfnamefont {D.}~\bibnamefont
			{Letscher}}, \ and\ \bibinfo {author} {\bibfnamefont {A.}~\bibnamefont
			{Zomorodian}},\ }\href {\doibase 10.1007/s00454-002-2885-2} {\bibfield
		{journal} {\bibinfo  {journal} {Discrete Comput. Geom.}\ }\textbf {\bibinfo
			{volume} {28}},\ \bibinfo {pages} {511} (\bibinfo {year} {2002})}\BibitemShut
	{NoStop}%
	\bibitem [{\citenamefont {Zomorodian}\ and\ \citenamefont
		{Carlsson}(2005)}]{zomorodian:2005}%
	\BibitemOpen
	\bibfield  {author} {\bibinfo {author} {\bibfnamefont {A.}~\bibnamefont
			{Zomorodian}}\ and\ \bibinfo {author} {\bibfnamefont {G.}~\bibnamefont
			{Carlsson}},\ }\href {\doibase 10.1007/s00454-004-1146-y} {\bibfield
		{journal} {\bibinfo  {journal} {Discrete Comput. Geom.}\ }\textbf {\bibinfo
			{volume} {33}},\ \bibinfo {pages} {249} (\bibinfo {year} {2005})}\BibitemShut
	{NoStop}%
	\bibitem [{\citenamefont {Chazal}\ \emph {et~al.}(2014)\citenamefont {Chazal},
		\citenamefont {de~Silva},\ and\ \citenamefont
		{Oudot}}]{chazal:stability:2014}%
	\BibitemOpen
	\bibfield  {author} {\bibinfo {author} {\bibfnamefont {F.}~\bibnamefont
			{Chazal}}, \bibinfo {author} {\bibfnamefont {V.}~\bibnamefont {de~Silva}}, \
		and\ \bibinfo {author} {\bibfnamefont {S.}~\bibnamefont {Oudot}},\ }\href
	{\doibase 10.1007/s10711-013-9937-z} {\bibfield  {journal} {\bibinfo
			{journal} {Geom. Dedicata}\ }\textbf {\bibinfo {volume} {173}},\ \bibinfo
		{pages} {193} (\bibinfo {year} {2014})}\BibitemShut {NoStop}%
	\bibitem [{\citenamefont {Reininghaus}\ \emph {et~al.}(2015)\citenamefont
		{Reininghaus}, \citenamefont {Huber}, \citenamefont {Bauer},\ and\
		\citenamefont {Kwitt}}]{reininghaus:mskernel:2015}%
	\BibitemOpen
	\bibfield  {author} {\bibinfo {author} {\bibfnamefont {J.}~\bibnamefont
			{Reininghaus}}, \bibinfo {author} {\bibfnamefont {S.}~\bibnamefont {Huber}},
		\bibinfo {author} {\bibfnamefont {U.}~\bibnamefont {Bauer}}, \ and\ \bibinfo
		{author} {\bibfnamefont {R.}~\bibnamefont {Kwitt}},\ }in\ \href {\doibase
		10.1109/CVPR.2015.7299106} {\emph {\bibinfo {booktitle} {Proc. 28th IEEE
				Conf. Computer Vision and Pattern Recognition (CVPR)}}}\ (\bibinfo {year}
	{2015})\BibitemShut {NoStop}%
	\bibitem [{\citenamefont {Carri{\`e}re}\ \emph {et~al.}(2017)\citenamefont
		{Carri{\`e}re}, \citenamefont {Cuturi},\ and\ \citenamefont
		{Oudot}}]{carrire:wskernel:2017}%
	\BibitemOpen
	\bibfield  {author} {\bibinfo {author} {\bibfnamefont {M.}~\bibnamefont
			{Carri{\`e}re}}, \bibinfo {author} {\bibfnamefont {M.}~\bibnamefont
			{Cuturi}}, \ and\ \bibinfo {author} {\bibfnamefont {S.}~\bibnamefont
			{Oudot}},\ }in\ \href {http://proceedings.mlr.press/v70/carriere17a.html}
	{\emph {\bibinfo {booktitle} {Proc. 34th Int. Conf. Machine Learning
				(ICML)}}},\ Vol.~\bibinfo {volume} {70}\ (\bibinfo {year} {2017})\BibitemShut
	{NoStop}%
	\bibitem [{\citenamefont {Gretton}\ \emph {et~al.}(2008)\citenamefont
		{Gretton}, \citenamefont {Fukumizu}, \citenamefont {Teo}, \citenamefont
		{Song}, \citenamefont {Sch\"{o}lkopf},\ and\ \citenamefont
		{Smola}}]{gretton:nips:2007}%
	\BibitemOpen
	\bibfield  {author} {\bibinfo {author} {\bibfnamefont {A.}~\bibnamefont
			{Gretton}}, \bibinfo {author} {\bibfnamefont {K.}~\bibnamefont {Fukumizu}},
		\bibinfo {author} {\bibfnamefont {C.~H.}\ \bibnamefont {Teo}}, \bibinfo
		{author} {\bibfnamefont {L.}~\bibnamefont {Song}}, \bibinfo {author}
		{\bibfnamefont {B.}~\bibnamefont {Sch\"{o}lkopf}}, \ and\ \bibinfo {author}
		{\bibfnamefont {A.~J.}\ \bibnamefont {Smola}},\ }in\ \href
	{http://papers.nips.cc/paper/3201-a-kernel-statistical-test-of-independence.pdf}
	{\emph {\bibinfo {booktitle} {Adv. Neural Inf. Process. Syst. 20 (NIPS)}}}\
	(\bibinfo {year} {2008})\BibitemShut {NoStop}%
	\bibitem [{\citenamefont {Tran}\ and\ \citenamefont {Hasegawa}()}]{gitsource}%
	\BibitemOpen
	\bibfield  {author} {\bibinfo {author} {\bibfnamefont {Q.~H.}\ \bibnamefont
			{Tran}}\ and\ \bibinfo {author} {\bibfnamefont {Y.}~\bibnamefont
			{Hasegawa}},\ }\href@noop {} {\enquote {\bibinfo {title} {Delay-variant
				embedding},}\ }\bibinfo {howpublished}
	{\url{https://github.com/OminiaVincit/delay-variant-embed}},\ \bibinfo {note}
	{\text{G}itHub Repository}\BibitemShut {NoStop}%
	\bibitem [{\citenamefont {Phillips}\ \emph {et~al.}(2017)\citenamefont
		{Phillips}, \citenamefont {Manning}, \citenamefont {Papalopulu},\ and\
		\citenamefont {Rattray}}]{phillips:hesplos:2017}%
	\BibitemOpen
	\bibfield  {author} {\bibinfo {author} {\bibfnamefont {N.~E.}\ \bibnamefont
			{Phillips}}, \bibinfo {author} {\bibfnamefont {C.}~\bibnamefont {Manning}},
		\bibinfo {author} {\bibfnamefont {N.}~\bibnamefont {Papalopulu}}, \ and\
		\bibinfo {author} {\bibfnamefont {M.}~\bibnamefont {Rattray}},\ }\href
	{\doibase 10.1371/journal.pcbi.1005479} {\bibfield  {journal} {\bibinfo
			{journal} {PLOS Comp. Biol.}\ }\textbf {\bibinfo {volume} {13}},\ \bibinfo
		{pages} {1} (\bibinfo {year} {2017})}\BibitemShut {NoStop}%
	\bibitem [{\citenamefont {Monk}(2003)}]{monk:hesmodel:2003}%
	\BibitemOpen
	\bibfield  {author} {\bibinfo {author} {\bibfnamefont {N.~A.}\ \bibnamefont
			{Monk}},\ }\href {\doibase 10.1016/S0960-9822(03)00494-9} {\bibfield
		{journal} {\bibinfo  {journal} {Curr. Biol.}\ }\textbf {\bibinfo {volume}
			{13}},\ \bibinfo {pages} {1409 } (\bibinfo {year} {2003})}\BibitemShut
	{NoStop}%
	\bibitem [{\citenamefont {Galla}(2009)}]{galla:fluctuation:2009}%
	\BibitemOpen
	\bibfield  {author} {\bibinfo {author} {\bibfnamefont {T.}~\bibnamefont
			{Galla}},\ }\href {\doibase 10.1103/PhysRevE.80.021909} {\bibfield  {journal}
		{\bibinfo  {journal} {Phys. Rev. E}\ }\textbf {\bibinfo {volume} {80}},\
		\bibinfo {pages} {021909} (\bibinfo {year} {2009})}\BibitemShut {NoStop}%
	\bibitem [{\citenamefont {Gillespie}(1977)}]{gillespie:exact:1977}%
	\BibitemOpen
	\bibfield  {author} {\bibinfo {author} {\bibfnamefont {D.~T.}\ \bibnamefont
			{Gillespie}},\ }\href {\doibase 10.1021/j100540a008} {\bibfield  {journal}
		{\bibinfo  {journal} {J. Phys. Chem.}\ }\textbf {\bibinfo {volume} {81}},\
		\bibinfo {pages} {2340} (\bibinfo {year} {1977})}\BibitemShut {NoStop}%
	\bibitem [{\citenamefont {Anderson}(2007)}]{anderson:gillespie:2007}%
	\BibitemOpen
	\bibfield  {author} {\bibinfo {author} {\bibfnamefont {D.~F.}\ \bibnamefont
			{Anderson}},\ }\href {\doibase 10.1063/1.2799998} {\bibfield  {journal}
		{\bibinfo  {journal} {J. Phys. Chem.}\ }\textbf {\bibinfo {volume} {127}},\
		\bibinfo {pages} {214107} (\bibinfo {year} {2007})}\BibitemShut {NoStop}%
	\bibitem [{\citenamefont {Brett}\ and\ \citenamefont
		{Galla}(2013)}]{brett:distributed:2013}%
	\BibitemOpen
	\bibfield  {author} {\bibinfo {author} {\bibfnamefont {T.}~\bibnamefont
			{Brett}}\ and\ \bibinfo {author} {\bibfnamefont {T.}~\bibnamefont {Galla}},\
	}\href {\doibase 10.1103/PhysRevLett.110.250601} {\bibfield  {journal}
		{\bibinfo  {journal} {Phys. Rev. Lett.}\ }\textbf {\bibinfo {volume} {110}},\
		\bibinfo {pages} {250601} (\bibinfo {year} {2013})}\BibitemShut {NoStop}%
	\bibitem [{\citenamefont {Bishop}(2006)}]{prml:bishop:2006}%
	\BibitemOpen
	\bibfield  {author} {\bibinfo {author} {\bibfnamefont {C.~M.}\ \bibnamefont
			{Bishop}},\ }\href
	{http://research.microsoft.com/en-us/um/people/cmbishop/prml/} {\emph
		{\bibinfo {title} {Pattern Recognition and Machine Learning}}}\ (\bibinfo
	{publisher} {Springer},\ \bibinfo {year} {2006})\BibitemShut {NoStop}%
	\bibitem [{\citenamefont {Chen}\ \emph {et~al.}(2015)\citenamefont {Chen},
		\citenamefont {Keogh}, \citenamefont {Hu}, \citenamefont {Begum},
		\citenamefont {Bagnall}, \citenamefont {Mueen},\ and\ \citenamefont
		{Batista}}]{UCRarchive}%
	\BibitemOpen
	\bibfield  {author} {\bibinfo {author} {\bibfnamefont {Y.}~\bibnamefont
			{Chen}}, \bibinfo {author} {\bibfnamefont {E.}~\bibnamefont {Keogh}},
		\bibinfo {author} {\bibfnamefont {B.}~\bibnamefont {Hu}}, \bibinfo {author}
		{\bibfnamefont {N.}~\bibnamefont {Begum}}, \bibinfo {author} {\bibfnamefont
			{A.}~\bibnamefont {Bagnall}}, \bibinfo {author} {\bibfnamefont
			{A.}~\bibnamefont {Mueen}}, \ and\ \bibinfo {author} {\bibfnamefont
			{G.}~\bibnamefont {Batista}},\ }\href@noop {} {\enquote {\bibinfo {title}
			{The \text{UCR Time Series Classification Archive}},}\ } (\bibinfo {year}
	{2015}),\ \bibinfo {note}
	{\url{www.cs.ucr.edu/~eamonn/time_series_data/}}\BibitemShut {NoStop}%
	\bibitem [{\citenamefont {Brown}\ \emph {et~al.}(2013)\citenamefont {Brown},
		\citenamefont {Yemini}, \citenamefont {Grundy}, \citenamefont {Jucikas},\
		and\ \citenamefont {Schafer}}]{brown:elegans:791}%
	\BibitemOpen
	\bibfield  {author} {\bibinfo {author} {\bibfnamefont {A.~E.~X.}\
			\bibnamefont {Brown}}, \bibinfo {author} {\bibfnamefont {E.~I.}\ \bibnamefont
			{Yemini}}, \bibinfo {author} {\bibfnamefont {L.~J.}\ \bibnamefont {Grundy}},
		\bibinfo {author} {\bibfnamefont {T.}~\bibnamefont {Jucikas}}, \ and\
		\bibinfo {author} {\bibfnamefont {W.~R.}\ \bibnamefont {Schafer}},\ }\href
	{\doibase 10.1073/pnas.1211447110} {\bibfield  {journal} {\bibinfo  {journal}
			{Proc. Natl. Acad. Sci. U.S.A.}\ }\textbf {\bibinfo {volume} {110}},\
		\bibinfo {pages} {791} (\bibinfo {year} {2013})}\BibitemShut {NoStop}%
	\bibitem [{\citenamefont {Yemini}\ \emph {et~al.}(2013)\citenamefont {Yemini},
		\citenamefont {Jucikas}, \citenamefont {Grundy}, \citenamefont {Brown},\ and\
		\citenamefont {Schafer}}]{yemini:database:2013}%
	\BibitemOpen
	\bibfield  {author} {\bibinfo {author} {\bibfnamefont {E.}~\bibnamefont
			{Yemini}}, \bibinfo {author} {\bibfnamefont {T.}~\bibnamefont {Jucikas}},
		\bibinfo {author} {\bibfnamefont {L.~J.}\ \bibnamefont {Grundy}}, \bibinfo
		{author} {\bibfnamefont {A.~E.}\ \bibnamefont {Brown}}, \ and\ \bibinfo
		{author} {\bibfnamefont {W.~R.}\ \bibnamefont {Schafer}},\ }\href {\doibase
		10.1038/nmeth.2560} {\bibfield  {journal} {\bibinfo  {journal} {Nat.
				Methods}\ }\textbf {\bibinfo {volume} {10}},\ \bibinfo {pages} {877}
		(\bibinfo {year} {2013})}\BibitemShut {NoStop}%
	\bibitem [{\citenamefont {Lines}\ and\ \citenamefont
		{Bagnall}(2014)}]{lines:ee:2014}%
	\BibitemOpen
	\bibfield  {author} {\bibinfo {author} {\bibfnamefont {J.}~\bibnamefont
			{Lines}}\ and\ \bibinfo {author} {\bibfnamefont {A.}~\bibnamefont
			{Bagnall}},\ }\href {\doibase 10.1007/s10618-014-0361-2} {\bibfield
		{journal} {\bibinfo  {journal} {Data Min. Knowl. Disc.}\ }\textbf {\bibinfo
			{volume} {29}},\ \bibinfo {pages} {565} (\bibinfo {year} {2014})}\BibitemShut
	{NoStop}%
	\bibitem [{\citenamefont {Grabocka}\ \emph {et~al.}(2014)\citenamefont
		{Grabocka}, \citenamefont {Schilling}, \citenamefont {Wistuba},\ and\
		\citenamefont {Schmidt-Thieme}}]{grabocka:LST:2014}%
	\BibitemOpen
	\bibfield  {author} {\bibinfo {author} {\bibfnamefont {J.}~\bibnamefont
			{Grabocka}}, \bibinfo {author} {\bibfnamefont {N.}~\bibnamefont {Schilling}},
		\bibinfo {author} {\bibfnamefont {M.}~\bibnamefont {Wistuba}}, \ and\
		\bibinfo {author} {\bibfnamefont {L.}~\bibnamefont {Schmidt-Thieme}},\ }in\
	\href {\doibase 10.1145/2623330.2623613} {\emph {\bibinfo {booktitle} {Proc.
				20th ACM SIGKDD Int. Conf. Knowledge Discovery and Data Mining}}}\ (\bibinfo
	{year} {2014})\BibitemShut {NoStop}%
	\bibitem [{\citenamefont {Bagnall}\ \emph {et~al.}(2016)\citenamefont
		{Bagnall}, \citenamefont {Lines}, \citenamefont {Bostrom}, \citenamefont
		{Large},\ and\ \citenamefont {Keogh}}]{bagnall:review:2016}%
	\BibitemOpen
	\bibfield  {author} {\bibinfo {author} {\bibfnamefont {A.}~\bibnamefont
			{Bagnall}}, \bibinfo {author} {\bibfnamefont {J.}~\bibnamefont {Lines}},
		\bibinfo {author} {\bibfnamefont {A.}~\bibnamefont {Bostrom}}, \bibinfo
		{author} {\bibfnamefont {J.}~\bibnamefont {Large}}, \ and\ \bibinfo {author}
		{\bibfnamefont {E.}~\bibnamefont {Keogh}},\ }\href {\doibase
		10.1007/s10618-016-0483-9} {\bibfield  {journal} {\bibinfo  {journal} {Data
				Min. Knowl. Disc.}\ }\textbf {\bibinfo {volume} {31}},\ \bibinfo {pages}
		{606} (\bibinfo {year} {2016})}\BibitemShut {NoStop}%
	\bibitem [{\citenamefont {Bagnall}\ \emph {et~al.}()\citenamefont {Bagnall},
		\citenamefont {Lines}, \citenamefont {Vickers},\ and\ \citenamefont
		{Keogh}}]{tscresults}%
	\BibitemOpen
	\bibfield  {author} {\bibinfo {author} {\bibfnamefont {A.}~\bibnamefont
			{Bagnall}}, \bibinfo {author} {\bibfnamefont {J.}~\bibnamefont {Lines}},
		\bibinfo {author} {\bibfnamefont {W.}~\bibnamefont {Vickers}}, \ and\
		\bibinfo {author} {\bibfnamefont {E.}~\bibnamefont {Keogh}},\ }\href@noop {}
	{\enquote {\bibinfo {title} {The \text{UEA \& UCR Time Series Classification
					Repository}},}\ }\bibinfo {note}
	{\url{www.timeseriesclassification.com}}\BibitemShut {NoStop}%
	\bibitem [{\citenamefont {Szegedy}\ \emph {et~al.}(2014)\citenamefont
		{Szegedy}, \citenamefont {Zaremba}, \citenamefont {Sutskever}, \citenamefont
		{Bruna}, \citenamefont {Erhan}, \citenamefont {Goodfellow},\ and\
		\citenamefont {Fergus}}]{szegedy:adversarial:2014}%
	\BibitemOpen
	\bibfield  {author} {\bibinfo {author} {\bibfnamefont {C.}~\bibnamefont
			{Szegedy}}, \bibinfo {author} {\bibfnamefont {W.}~\bibnamefont {Zaremba}},
		\bibinfo {author} {\bibfnamefont {I.}~\bibnamefont {Sutskever}}, \bibinfo
		{author} {\bibfnamefont {J.}~\bibnamefont {Bruna}}, \bibinfo {author}
		{\bibfnamefont {D.}~\bibnamefont {Erhan}}, \bibinfo {author} {\bibfnamefont
			{I.}~\bibnamefont {Goodfellow}}, \ and\ \bibinfo {author} {\bibfnamefont
			{R.}~\bibnamefont {Fergus}},\ }in\ \href {http://arxiv.org/abs/1312.6199}
	{\emph {\bibinfo {booktitle} {Proc. 2nd Int. Conf. Learning Representations
				(ICLR)}}}\ (\bibinfo {year} {2014})\BibitemShut {NoStop}%
\end{thebibliography}
\end{document}